\documentclass[fleqn,usenatbib]{mnras}
\usepackage{newtxtext,newtxmath}
\usepackage[T1]{fontenc}

\DeclareRobustCommand{\VAN}[3]{#2}
\let\VANthebibliography\thebibliography
\def\thebibliography{\DeclareRobustCommand{\VAN}[3]{##3}\VANthebibliography}

\usepackage{graphicx}
\usepackage{amsmath}
\usepackage{multirow}

\title[North-south asymmetry of ALFALFA H\,{\textsc{i}}\,WF]{The north-south asymmetry of the ALFALFA H\,{\textsc{i}} velocity width function}

\author[R. A. N. Brooks et al.]{
Richard A. N. Brooks,$^{1,2,3}$\thanks{E-mail: richard.brooks.22@ucl.ac.uk}
Kyle A. Oman,$^{2,3}$
Carlos. S. Frenk $^{2,3}$
\\
$^{1}$Department of Physics and Astronomy, University College London, London WC1E 6BT, UK\\
$^{2}$Institute for Computational Cosmology, Durham University, South Road, Durham DH1 3LE, United Kingdom\\
$^{3}$Department of Physics, Durham University, South Road, Durham DH1 3LE, UK\\
}

\date{Accepted XXX. Received YYY; in original form ZZZ}

\pubyear{\the\year}

\begin{document}
\label{firstpage}
\pagerange{\pageref{firstpage}--\pageref{lastpage}}
\maketitle

\begin{abstract}
The number density of extragalactic {21-cm} radio sources as a function of their spectral line-widths -- the H\,\textsc{i} width function (H\,\textsc{i}\,WF) -- is a sensitive tracer of the dark matter halo mass function (HMF). The $\Lambda$ cold dark matter model predicts that the HMF should be identical everywhere provided it is sampled in sufficiently large volumes, implying that the same should be true of the H\,\textsc{i}\,WF. The ALFALFA {21-cm} survey measured the H\,\textsc{i}\,WF in northern and southern Galactic fields and found a systematically higher number density in the north. At face value, this is in tension with theoretical predictions. We use the \emph{Sibelius-DARK} N-body simulation and the semi-analytical galaxy formation model \emph{GALFORM} to create a mock ALFALFA survey. We find that the offset in number density has two origins: the sensitivity of the survey is different in the two fields, which has not been correctly accounted for in previous measurements; and the $1/V_{\mathrm{eff}}$ algorithm used for completeness corrections does not fully account for biases arising from spatial clustering in the galaxy distribution. The latter is primarily driven by a foreground overdensity in the northern field within $30\,\mathrm{Mpc}$, but more distant structure also plays a role. We provide updated measurements of the ALFALFA H\,\textsc{i}\,WF (and H\,\textsc{i}\,MF) correcting for the variations in survey sensitivity. Only when systematic effects such as these are understood and corrected for can cosmological models be tested against the H\,\textsc{i}\,WF.

\end{abstract}


\section{Introduction}\label{sec:Introduction} 

The standard Lambda Cold Dark Matter ($\Lambda$CDM) cosmological model predicts that the number density of self-bound dark matter haloes as a function of mass -- the Halo Mass Function \citep[HMF, ][]{1988ApJ...327..507F} -- is well-approximated by a power law with slope of $\phi(M) \propto M^{-1.9}$ over almost 20 orders of magnitude below the scale of the largest collapsed structures today \citep{2020Natur.585...39W}. The low-mass end of the HMF is sensitive to the the power spectrum of density fluctuations in the early universe. For instance, for thermal relic dark matter particles that are lighter than those assumed in $\Lambda$CDM models \citep[e.g., warm dark matter;][]{2001ApJ...559..516A,Bode_2001}, there is a cut-off in the power spectrum at smaller scales. The higher velocities of lighter thermal relics naturally suppress the formation of low-mass haloes because free-streaming effects erase the density perturbations that could seed them. Measuring the HMF therefore offers an opportunity to constrain the particle nature of dark matter. However, the HMF is not directly measurable and hence indirect measurements of the HMF must be made instead. One option is to study the abundances of galaxies as a function of the kinematics of visible tracers orbiting within their dark matter haloes.

There is an apparent tension between the number of low-mass dark matter haloes predicted by $\Lambda$CDM dark matter-only simulations in which galaxies are expected to form, and the number of galaxies with kinematics compatible with inhabiting them measured from observation. This problem was first identified for satellite galaxies of the Milky Way \citep[‘too-big-to-fail',][]{2011MNRAS.415L..40B}, and subsequently resolved by \citet{2016MNRAS.457.1931S} who showed that the inclusion of baryons in the simulations changes the theoretical predictions; the inevitable ejection of baryons from haloes at early times (due to supernova ‘feedback') reduces the growth rate of haloes such that, at the present time, a halo in the full-physics simulation is about $10$~per~cent less massive than its counterpart in the dark matter-only simulation. A similar issue applying to field galaxies was identified by \citet{2015A&A...574A.113P}, and it is so far unclear whether this issue is similarly resolved. In this instance, dwarf galaxies should be hosted by dark matter haloes that are significantly more massive than those implied by measuring the kinematics within the galaxies. If instead lower mass dark matter halos host dwarf galaxies, observational surveys should measure a much higher number density of galaxies.

The kinematics of observable tracers in a galaxy are linked to the HMF because the maximum circular velocity of a halo, $v_\mathrm{max}$, is correlated to its mass \citep{1997ApJ...490..493N}. Connecting a kinematic tracer to $v_\mathrm{max}$ often requires additional modelling, e.g. in the case of the {21-cm} spectral line width. The {21-cm} velocity spectrum is the H\,\textsc{i} mass-weighted line-of-sight velocity distribution. The {21-cm} line-width can be parameterised as the full width at half maximum of the spectrum, $w_{50}$. If a dark matter halo contains a sufficiently extended H\,\textsc{i} disc, the maximum circular velocity of the dark matter halo is approximately $v_{\mathrm{max}} \sim w_{50}/2\sin(i)$, where $i$ is the inclination angle to the line of sight. However, there are considerations which need to be made when computing $v_\mathrm{max}$ using this approach. In many cases galaxies will have gas discs that are not sufficiently extended to reach the flat part of the velocity profile \citep{2015MNRAS.453.2133B, 2016MNRAS.463.4052P, 2016MNRAS.455.3841B, 2016MNRAS.463L..69M, 2017ApJ...850...97B}. Additionally, the inclination angle is problematic because line-width surveys usually do not spatially resolve the gas structure and therefore rely on optical counterparts of radio sources to get an estimate for $i$ \citep[e.g., ][]{2010MNRAS.403.1969Z}. There are many other difficulties faced when obtaining $v_\mathrm{max}$ using line-width measurements, such as: the gas orbits may not be circular \citep{2016MNRAS.455.3841B}; the gas disc may not lie in a single plane \citep{2016MNRAS.463.4052P}; emission may be mistakenly taken to overlap with a neighbouring source \citep{2015MNRAS.449.1856J, 2019MNRAS.488.5898C}; the gas disc may be partially supported by turbulent or thermal pressure \citep{2015MNRAS.453.2133B, 2016MNRAS.463.4052P}; etc. Instead of attempting to infer $v_\mathrm{max}$ from line-width measurements, a potentially more straightforward technique is to predict the number density of extragalactic {21-cm} sources as a function of $w_{50}$ -- the H\,\textsc{i} width function (H\,\textsc{i}\,WF) -- and then compare with observational measurements. Comparison of the H\,\textsc{i}\,WF from theory to an observational equivalent thus provides an alternative avenue to investigate dark matter than to try and infer a HMF from an observed H\,\textsc{i}\,WF.

For the Arecibo Legacy Fast ALFA\footnote{Arecibo L-band Feed Array} \citep[ALFALFA;][]{2005AJ....130.2598G} survey, the H\,\textsc{i}\,WF and H\,\textsc{i} mass function (H\,\textsc{i}\,MF) are measured simultaneously as orthogonal integrations of the two-dimensional H\,\textsc{i} mass-width function. \citet{2018MNRAS.477....2J} reported that the H\,\textsc{i}\,MF has significantly different global shapes in the ‘spring' (northern Galactic hemisphere) and ‘fall' (southern Galactic hemisphere) fields of the survey. The low-mass slope is significantly shallower the in fall field. The large-scale environmental dependence of the low-mass slope in the H\,\textsc{i}\,MF was tentatively attributed to the presence of the Virgo cluster in the foreground of one half of the survey and a deep void in the other \citep{2018MNRAS.477....2J}. The H\,\textsc{i}\,WF, on the other hand, has been reported to have a similar shape in both survey fields. \citet{2022MNRAS.509.3268O} tentatively attributed this similarity in shape to possible environmental effects affecting the shape of the H\,\textsc{i}\,MF, but leaving the shape of the H\,\textsc{i}\,WF largely unchanged. However, they made no attempt to explain the different overall normalisation of the H\,\textsc{i}\,MFs and H\,\textsc{i}\,WFs in the two regions.

The first measurements of the H\,\textsc{i}\,WF came almost simultaneously from two different {21-cm} surveys. \cite{2009ApJ...700.1779Z} used an early release of the ALFALFA survey with only $6$~per~cent of the final data available. Meanwhile, \cite{2010MNRAS.403.1969Z} used the H\,\textsc{i} Parkes All-Sky Survey (H\,\textsc{i}\,PASS). Both of these measurements revealed that the $\Lambda$CDM model apparently overpredicts the abundance of sources at the low velocity-width end. The same overabundance problem persists in follow-up work after subsequent ALFALFA data releases \citep{2011ApJ...739...38P, 2018MNRAS.477....2J}, but can be explained (at least to leading order) by the systematic reduction of the total mass an abundance of structures below $v_\mathrm{max}\sim 100\,\mathrm{km}\,\mathrm{s}^{-1}$ by gas pressure, reionization, supernova feedback, stripping, and truncated accretion \citep[][see also \citealp{2016MNRAS.463L..69M,2016A&A...591A..58P,2017ApJ...850...97B,2019MNRAS.482.5606D}]{2013MNRAS.431.1366S}.

\cite{2009ApJ...700.1779Z} and \cite{2010MNRAS.403.1969Z} both measured differences in the normalisation of the H\,\textsc{i}\,WF in different subsets of the ALFALFA and H\,\textsc{i}\,PASS surveys, respectively, although their shapes are statistically consistent with being identical. It has been variously speculated that what drives the difference in the normalisation of the H\,\textsc{i}\,WF between the ALFALFA spring and fall fields is a combination of: sample variance, distance modelling, the adopted completeness limit, etc. \citep[see,][for further discussion]{2022MNRAS.509.3268O}.

In this work we use the \emph{Sibelius-DARK} N-body simulation \citep{2022MNRAS.512.5823M} that reproduces the local structure of the Universe on scales larger than $\approx4\,\mathrm{Mpc}$ populated with galaxies using the \emph{GALFORM} semi-analytical model \citep{2016MNRAS.462.3854L} to create mock surveys similar to the ALFALFA survey. The nature of the \emph{Sibelius} simulations allows us to investigate the influence of the spatial clustering of galaxies along the line of sight, as well as possible environmental effects on the global shape of the H\,\textsc{i}\,WF. Previous work has only been able to comment speculatively on the origin of the differences in the H\,\textsc{i}\,WF between the two fields surveyed by ALFALFA \citep{2018MNRAS.477....2J, 2022MNRAS.509.3268O}. Our mock surveys provide a suitable footing for an investigation into survey systematics that may be responsible for driving the asymmetry between the spring and fall fields; our approach enables us to provide the first quantitative estimates for the magnitude of these effects and assess whether they can explain the observed asymmetry. 

The plan of this paper is as follows. In Sec.~\ref{sec:Methods} we outline the process we use to construct and analyse mock ALFALFA-like {21-cm} surveys. In Sec.~\ref{sec:results_HIWF} we present the ALFALFA and \emph{Sibelius-DARK} plus \emph{GALFORM} H\,\textsc{i}\,WFs under various conditions and assumptions. In Sec.~\ref{sec:Discussion} we comment on possible mitigations for variations in sensitivity for future surveys, and on the possible origins of qualitative differences between our mock H\,\textsc{i}\,MFs and H\,\textsc{i}\,WFs and those measured using ALFALFA. We summarise in Sec.~\ref{sec:ConstraintDM_conc}.

\section{Methods}\label{sec:Methods} 

In Sec.~\ref{sec:ALFALFA_Survey} we summarise the defining properties of the ALFALFA survey, detail how extragalactic sources were identified and show how the final survey catalogue is obtained. Next, in Sec.~\ref{sec:sibelius} we provide an overview of the N-body \emph{Sibelius-DARK} simulation and assess the assumption of identical HMFs in both fields. In Sec.~\ref{sec:galform} we give an overview of the \emph{GALFORM} semi-analytical model. In Sec.~\ref{sec:mock_survey} we explain how our \emph{GALFORM} mock {21-cm} survey is created. Finally, in Sec.~\ref{sec:veff_estimator}, we outline the $1/V_\mathrm{{eff}}$ statistical estimator used to correct the H\,\textsc{i}\,WF for observational incompleteness.

\subsection{The ALFALFA survey} \label{sec:ALFALFA_Survey} 

\begin{figure*}
    \centering
    \includegraphics[width=0.9\linewidth]{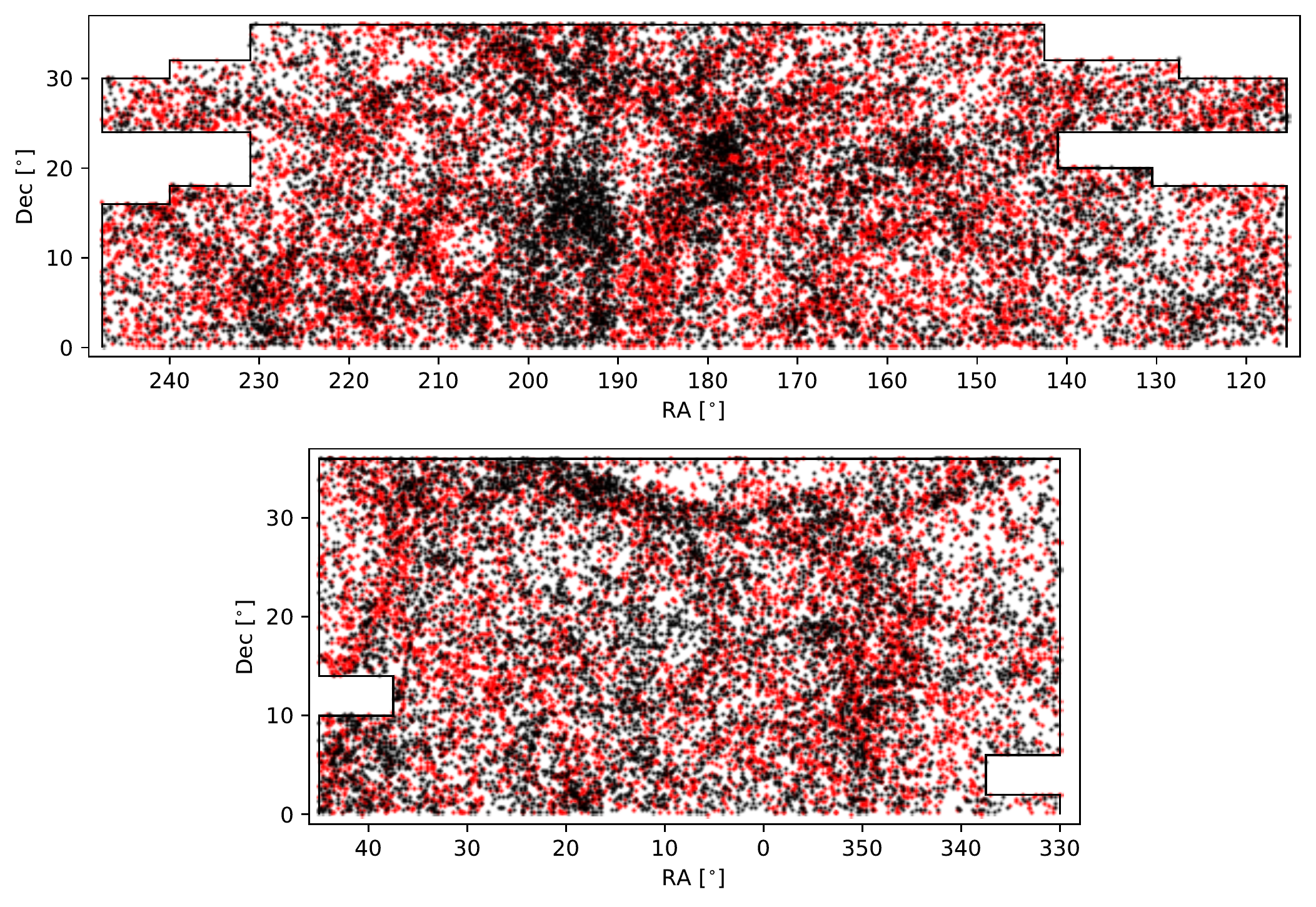}\\
    \includegraphics[width=0.85\linewidth]{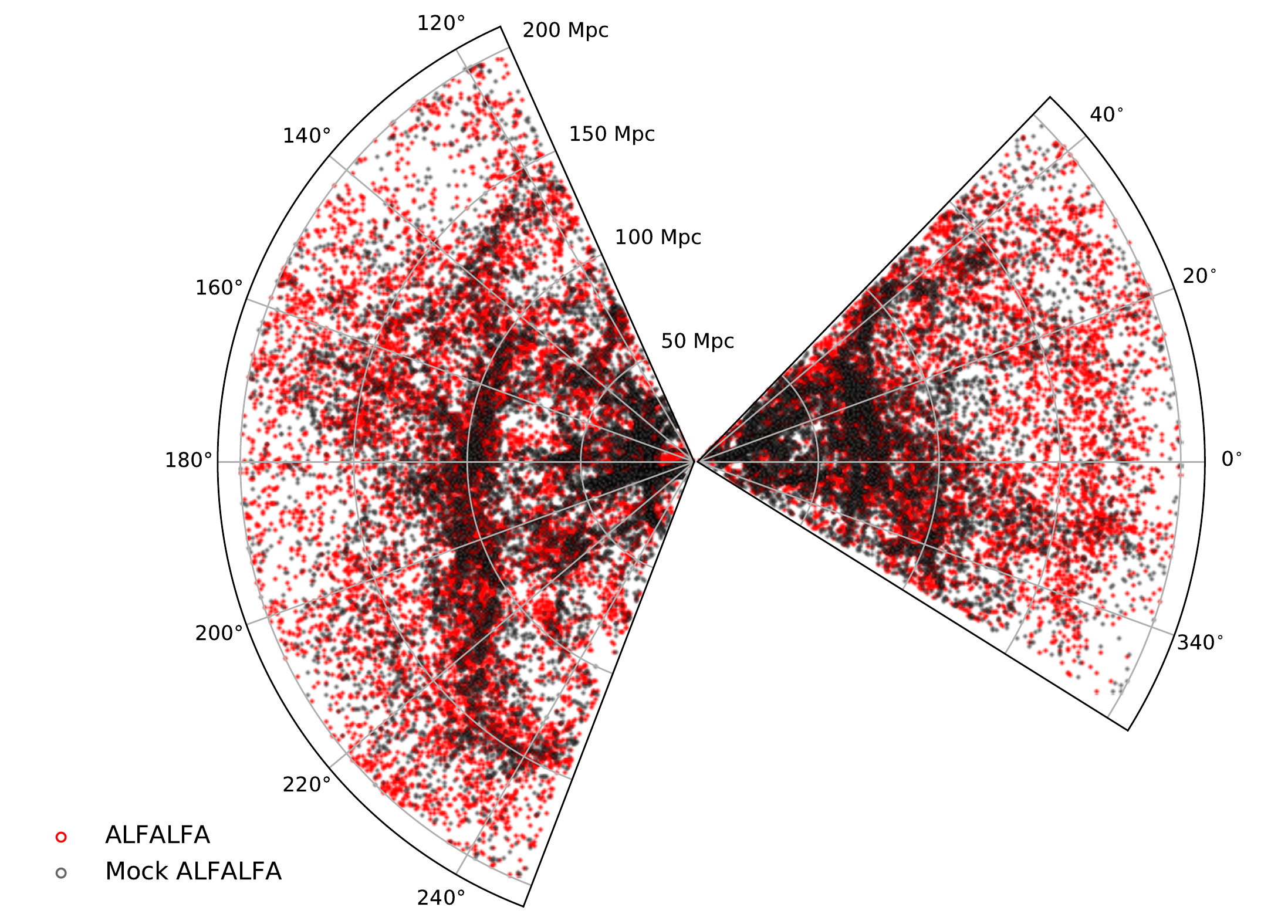}
    \caption{\emph{Top panel:} Distribution of $\alpha.100$ ALFALFA sources (red points) within the footprint on the sky of the spring field. Black points show sources contained in our ALFALFA-like mock survey catalogue constructed from the \emph{Sibelius-DARK} + \emph{GALFORM} simulation. \emph{Middle panel:} Same as top panel, but for the fall field. \emph{Bottom panel:} Cone diagram for the survey catalogues over all declinations as a function of distance from the Milky~Way and right ascension on the sky.}
    \label{fig:footprints_coneplots_subplots}
\end{figure*}

The ALFALFA survey \citep{2005AJ....130.2598G} mapped $\sim 7,000 \deg^2$ of the sky visible from Arecibo at {21-cm} wavelengths out to $\sim250\,\mathrm{Mpc}$, or $cz \leq 18,000\,\mathrm{km}\,\mathrm{s}^{-1}$. ALFALFA was specifically designed to investigate the faint end of the H\,\textsc{i}\,MF in the Local Universe. The survey was completed in 2012, and is composed of two separate fields on the sky; one in the northern Galactic hemisphere, visible during the spring, and the other in the southern Galactic hemisphere, visible during the autumn. By convention, these fields are labeled ‘spring' and ‘fall', respectively \citep{2018MNRAS.477....2J}. 

Extragalactic sources in the ALFALFA survey were identified using a matched-filtering technique \citep{2007AJ....133.2087S}, supplemented with some sources identified from direct inspection of the raw data cubes. These identified sources were subsequently manually checked to confirm or reject each individual detection, and to assign optical counterparts to detections where possible. The final product is the $\alpha.100$ extragalactic source catalogue described in \cite{2018ApJ...861...49H}. This catalogue lists the coordinates (for the H\,\textsc{i} and associated optical sources), redshifts, {21-cm} line flux densities, {21-cm} line widths, distances\footnote{\citet{2022arXiv221208728B} provides independent confirmation that the catalogue distances derived using the \citet{2005PhDT.........2M} flow model are accurate.}, signal-to-noise ratios and H\,\textsc{i} masses of sources, and their uncertainties where relevant. The H\,\textsc{i} mass, $M_{\mathrm{{HI}}}$, of a galaxy is determined as usual from the flux and distance as:
\begin{equation}\label{mHI_fromS21+distance}
    \frac{M_{\mathrm{HI}}}{\mathrm{M}_{\odot}} = 2.36~\times10^{5}\, \left(\frac{D}{\mathrm{Mpc}}\right)^{2}\,\frac{S_{21}}{\mathrm{Jy}\,\mathrm{km}\,\mathrm{s}^{-1}}.
\end{equation}
We define a selection of $\alpha.100$ sources from which we measure the H\,\textsc{i}\,WF in this work in a similar way to \citet{2022MNRAS.509.3268O}. This includes the choice of only ‘Code 1' (i.e., $\mathrm{S}/\mathrm{N} > 6.5$) sources whose (RA, Dec) coordinates fall in the survey footprint\footnote{See their tables D1-D4. We adopt the fiducial, not the ‘strict' footprint throughout this work.} \citep{2018MNRAS.477....2J}. Instead of the recessional velocity cut $v_{\mathrm{rec}}\,\leq 15,000\,\mathrm{km}\,\mathrm{s}^{-1}$ used in \citet{2022MNRAS.509.3268O}, we impose a distance cut $\mathrm{d_{mw} \leq 200\,\mathrm{Mpc}}$ in order to facilitate comparison with the \emph{Sibelius-DARK} simulation which is contaminated by low-resolution particles from outside of the zoom-in region \citep{2022MNRAS.512.5823M} beyond this distance. Only sources above the $50$~per~cent completeness limit (CL) of the survey are selected. The determination of the CL for the ALFALFA survey is described in \cite{2011AJ....142..170H}. For a flux-limited sample drawn from a uniformly-distributed population of galaxies, number counts as a function of flux are expected to follow a power law with exponent $-3/2$. Deviation from this form indicates the onset of incompleteness in the survey. There are $20,857$ sources above the $50$~per~cent CL in the $\alpha.100$ catalogue, of which $13,006$ are in the spring field and $7,851$ are in the fall field (see Table~\ref{table:galaxy_characteristics}). Fig.~\ref{fig:footprints_coneplots_subplots} visualises the $\alpha.100$ sources within the survey footprint on the sky (upper panels), and in a cone diagram over all declinations (lower panel). \citet{2022MNRAS.509.3268O} used the global $50$~per~cent CL that they derived throughout their analysis. We adopt the same CL in some contexts, but also make use of the $50$~per~cent CL derived separately from each of the two survey fields. The fall CL is slightly shallower than the global CL, by $0.011\,\mathrm{dex}$, while the spring CL is slightly deeper, by $0.009\,\mathrm{dex}$, for a net difference of $0.02\,\mathrm{dex}$. Explcitly:
\begin{equation} \label{50_CL_S}
    \mathrm{Spring}:\,\log_{10}\left(\frac{S_{21,50\%}}{\mathrm{Jy}\,\mathrm{km}\,\mathrm{s}^{-1}}\right) =
    \begin{cases}
      0.5\,W - 1.179 & \text{$W < 2.5$},\\
      W - 2.429 & \text{$W \geq 2.5$}\\
    \end{cases};
\end{equation}

\begin{equation} \label{50_CL_F}
    \mathrm{Fall}:\,\log_{10}\left(\frac{S_{21,50\%}}{\mathrm{Jy}\,\mathrm{km}\,\mathrm{s}^{-1}}\right) =
    \begin{cases}
      0.5\,W - 1.159 & \text{$W < 2.5$},\\
      W - 2.409 & \text{$W \geq 2.5$}\\
    \end{cases},
\end{equation}
where $W = \log_{10}(w_{50}\,/\mathrm{km}\,\mathrm{s}^{-1})$. The CLs at other completeness levels ($25$~per~cent, $90$~per~cent) are given in Appendix~\ref{app:CLs}.

\subsection{\emph{Sibelius-DARK}} \label{sec:sibelius} 

The ‘Simulations Beyond The Local Universe' \citep[\emph{Sibelius};][]{2022MNRAS.509.1432S} project aims to connect the Local Group (LG) with its cosmological environment. \emph{Sibelius} simulations use $\Lambda$CDM initial conditions that are constrained such that the large-scale structure is accurately reproduced, e.g. well-known galaxy clusters such as Virgo, Coma and Perseus are embedded within the correct large-scale cosmic web, and have appropriate masses. The initial conditions are generated using the \emph{BORG} algorithm \citep[‘Bayesian Origin Reconstruction from Galaxies’: ][]{2013MNRAS.432..894J} which derives initial conditions using Bayesian inference through forward modelling with Hamiltonian Monte Carlo methods, inferred in this instance from the 2M++ galaxy sample \citep{2011MNRAS.416.2840L}. The inference algorithm is fully probabilistic in the sense that it turns the task of reproducing the present non-linear galaxy distribution into a statistical initial conditions problem.

\begin{figure}
    \centering
	\includegraphics[width=\linewidth]{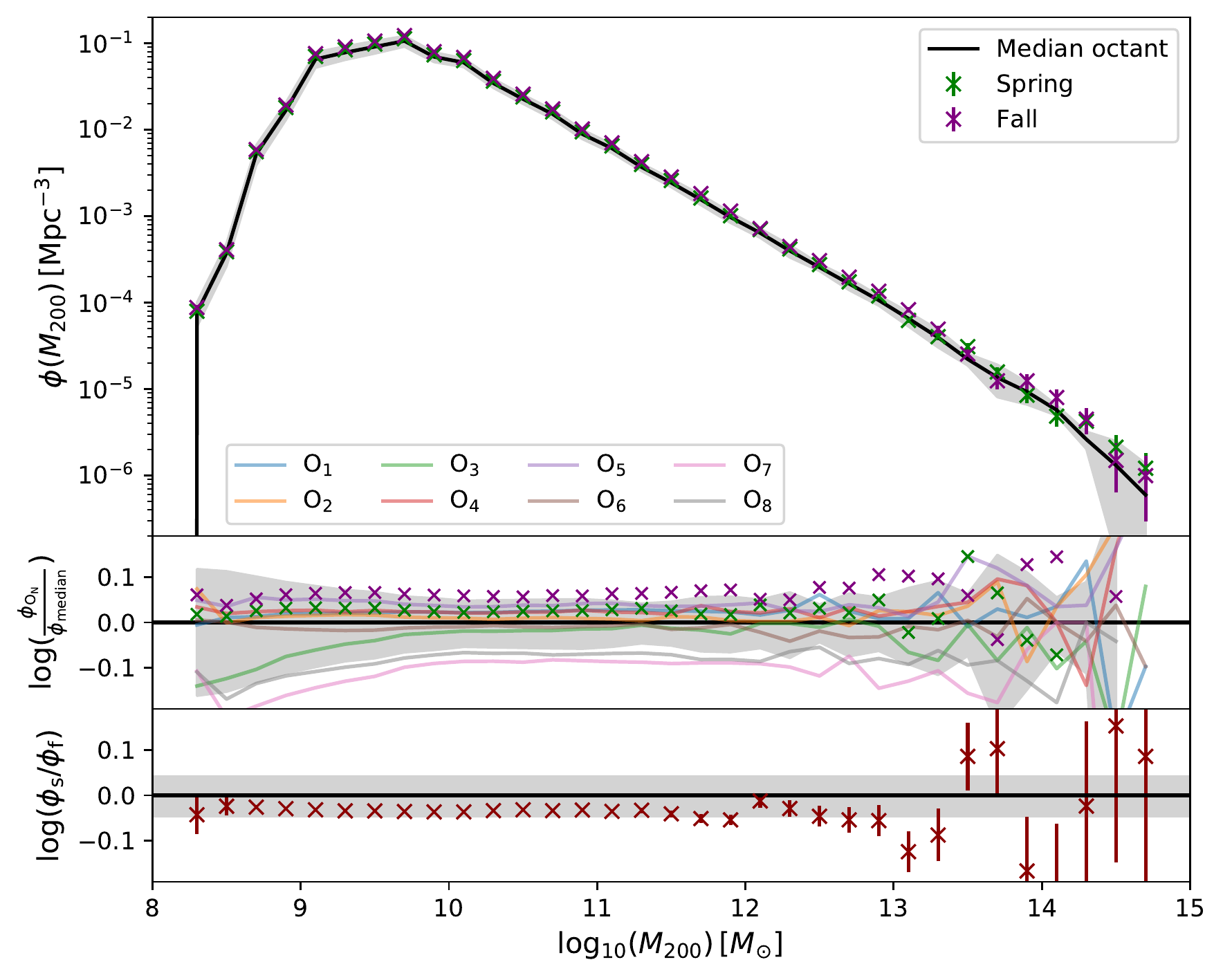}
    \caption{\textit{Top panel}: The halo mass function (HMF) for the spring (green crosses with $1\sigma$ Poisson uncertainties) and fall (purple crosses with $1\sigma$ uncertainties) ALFALFA fields from the $z=0$ \emph{Sibelius-Dark} plus \emph{GALFORM} output. Additionally, the median HMF for the eight separate octants with its inter-quartile scatter is shown (black crosses). The octants are denoted by $\mathrm{O_{N}}$ ($N = 1-8$); details of the octants are given in Appendix~\ref{appendix:octants}. \textit{Middle panel}: The ratio of the $\mathrm{N^{th}}$ octant's HMF to the median HMF across octants. Each coloured line represents one of the octants HMF ratios. The shaded region shows the inter-quartile scatter about the median. \textit{Bottom panel}: The ratio of the spring to fall HMF (dark-red crosses with $1\sigma$ uncertainties). The shaded region indicates a region of $\pm10$~per~cent differences in the HMF.}
    \label{fig:X}
\end{figure}

The first simulation from the \emph{Sibelius} project is \emph{Sibelius-DARK} \citep{2022MNRAS.512.5823M}, a realisation of a volume constrained within $200\,\mathrm{Mpc}$ of the Milky Way. Its volume makes it ideal to compare with the ALFALFA survey, which detects galaxies out to slightly beyond $200\,\mathrm{Mpc}$.

The simulation assumes a flat CDM cosmology with parameters from the \citet{2014A&A...571A..16P}: $\Omega_{\Lambda} = 0.693$, $\Omega_{m} = 0.307$, $\Omega_{b} = 0.04825$, $\sigma_8 = 0.8288$,
$n_s = 0.9611$, and $H_0 = 67.77 \,\mathrm{km}\,\mathrm{s}^{-1}\,\mathrm{Mpc}^{-3}$. We note that \citet{2018ApJ...861...49H} assume a slightly different cosmology that will cause $\sim10$~per~cent differences in the normalisation of the ALFALFA and our mock H\,{\textsc{i}}\,MFs and H\,{\textsc{i}}\,WFs. This is driven by the assumed value of the reduced Hubble constant, $h$, which when incorporated into the units of the H\,{\textsc{i}}\,MF and H\,{\textsc{i}}\,WF results in a different normalisation. Below, we restrict ourselves to qualitative comparisons between ALFALFA and \emph{GALFORM}, so this small quantitative difference does not influence our conclusions.

\subsubsection{The halo mass function of Sibelius-DARK}

In the top panel of Fig.~\ref{fig:X} we show the HMF for the spring and fall fields as well as for the median across eight octants of the \emph{Sibelius-DARK} sky (the spring and fall regions have volumes equivalent to $80$ and $50$~per~cent of an octant, respectively; all octants and their properties are detailed in Appendix~\ref{appendix:octants}). The middle panel of Fig.~\ref{fig:X} shows the ratio of the $\mathrm{N^{th}}$ octant's HMF to that of the median. The scatter in the HMF across the octants on the sky is $\lesssim20$~per~cent around the median.

 We quantify how closely we should expect the HMFs in regions of these volumes to agree as follows. \citet{2003ApJ...584..702H} provide a relation for the fractional variance $\sigma_n$ in the number density of haloes above a given mass threshold $M_\mathrm{th}$ in their fig.~2. We can approximate the low-mass end of their sample variance relation (sample variance dominates over shot noise in our mass range of interest) as $\sigma_n\propto k \left(\frac{M_\mathrm{th}}{10^{13}\,\mathrm{M}_\odot}\right)^{0.15}$, where $k$ depends on the survey volume. Assuming an `effective' survey radius of $111\,\mathrm{Mpc}$ -- the radius of a sphere enclosing the same volume as surveyed in the entire ALFALFA survey -- we obtain $k\sim0.116$. We therefore expect the mass function between $10^9$ and $10^{12}\,\mathrm{M}_\odot$ sampled in a volume equivalent to the ALFALFA survey to scatter by about $5$~per~cent around the cosmic mean HMF in an equivalent volume (or a factor of $\sqrt(2)$ more for a volume equivalent to half of the survey). Strictly speaking this calculation applies to cumulative mass functions, but the weak dependence on mass across our range of interest means that it also provides a reasonable order-of-magnitude estimate for the differential mass function.

Given that the fall field has a HMF that is about $8$~per~cent overdense with respect to the spring field, shown in the lower panel of Fig.~\ref{fig:X}, and that both fields are overdense with respect to the median (Fig.~\ref{fig:X}, middle panel), by about 16 and 6~per~cent, respectively are therefore consistent with the expectation for a $\Lambda$CDM cosmology.

Further reassurance is provided by the fact that the normalisations of the HMFs in the various octants and in the spring and fall survey regions are broadly consistent with the relative galaxy densities of the northern and southern hemispheres reported for \emph{Sibelius-DARK} by \citet[][sec~3.1.2 and fig.~8]{2022MNRAS.512.5823M} and similarly for the \emph{Simulating the LOcal Web} constrained realisation \citet[][]{2023arXiv230210960D}. The former highlighting that this difference is entirely consistent (within $1-2\,\sigma$) with the expectation for a $\Lambda$CDM cosmology. 


\subsection{\emph{GALFORM}}\label{sec:galform} 

Since the \emph{Sibelius-DARK} simulation is a N-body simulation, we model the evolution of the galaxy population using the \emph{GALFORM} semi-analytical model of galaxy formation.

The \emph{GALFORM} semi-analytical model calculates the evolution of galaxies in hierarchical theories of structure formation. The processes governing galaxy formation and evolution are modelled as sets of coupled non-linear differential equations \citep{1991ApJ...379...52W}. Since the first \emph{GALFORM} models \citep{1994MNRAS.271..781C, 2000MNRAS.319..168C}, there have been numerous changes and improvements \citep[e.g.,][]{2000ApJ...542..710G, 2000MNRAS.311..793B, 2005MNRAS.356.1191B}. We model the \emph{Sibelius-DARK} simulation galaxy population using the \citet{2016MNRAS.462.3854L} variant of \emph{GALFORM}. This is the same variant as used by \citet{2022MNRAS.512.5823M}, who present a detailed investigation into the resulting galaxy population (e.g., luminosity function in their fig. 5, stellar mass functions in their fig, 6, etc.). This \emph{GALFORM} variant incorporates different initial mass functions for quiescent star formation versus for starbursts, black hole formation, feedback from supernovae and from active galactic nuclei that suppresses gas cooling in massive haloes, and a new empirical star formation law in galaxy discs based on molecular gas content. A more accurate treatment of dynamical friction acting on satellite galaxies is also introduced, as well as an updated stellar population model.

\subsection{Mock {21-cm} survey}\label{sec:mock_survey} 

\begin{table*}
\centering
\caption{Characteristics of the $\alpha.100$ ALFALFA and mock \emph{Sibelius-DARK} + \emph{GALFORM} survey catalogues. Additionally shown is the information of the GALFORM galaxy population before any selection criteria are applied. The assumed distance limit is $200\,\mathrm{Mpc}$. The area of the spring and fall fields are $1.240\,\mathrm{sr}$ and $0.752\,\mathrm{sr}$, respectively.}
\label{tab:my_label}
\begin{tabular}{|c|c|c|c|}
 \hline
 \multicolumn{4}{|c|}{Characteristics of Galaxy Catalogues - Separate Completeness Limit} \\
 \hline
 & ALFALFA Survey & GALFORM Survey & GALFORM before Completeness Limit\\
 \hline
 Number of Spring Sources, $N_{s}$& 13006 & 12408 & 2074665\\
 Number of Fall Sources, $N_{f}$ & 7851 & 8623 & 1367320\\
 Number density of Spring Sources, $n_{s}$  & 0.0040 $\mathrm{Mpc^{-3}}$ & 0.0038 $\mathrm{Mpc^{-3}}$
 & 0.63 $\mathrm{Mpc^{-3}}$\\
 Number density of Fall Sources, $n_{f}$ & 0.0039 $\mathrm{Mpc^{-3}}$ & 0.0044 $\mathrm{Mpc^{-3}}$& 0.69 $\mathrm{Mpc^{-3}}$\\
 \hline
 
 \label{table:galaxy_characteristics}
\end{tabular}
\end{table*}

\begin{figure*}
    \centering
	\includegraphics[width=\linewidth]{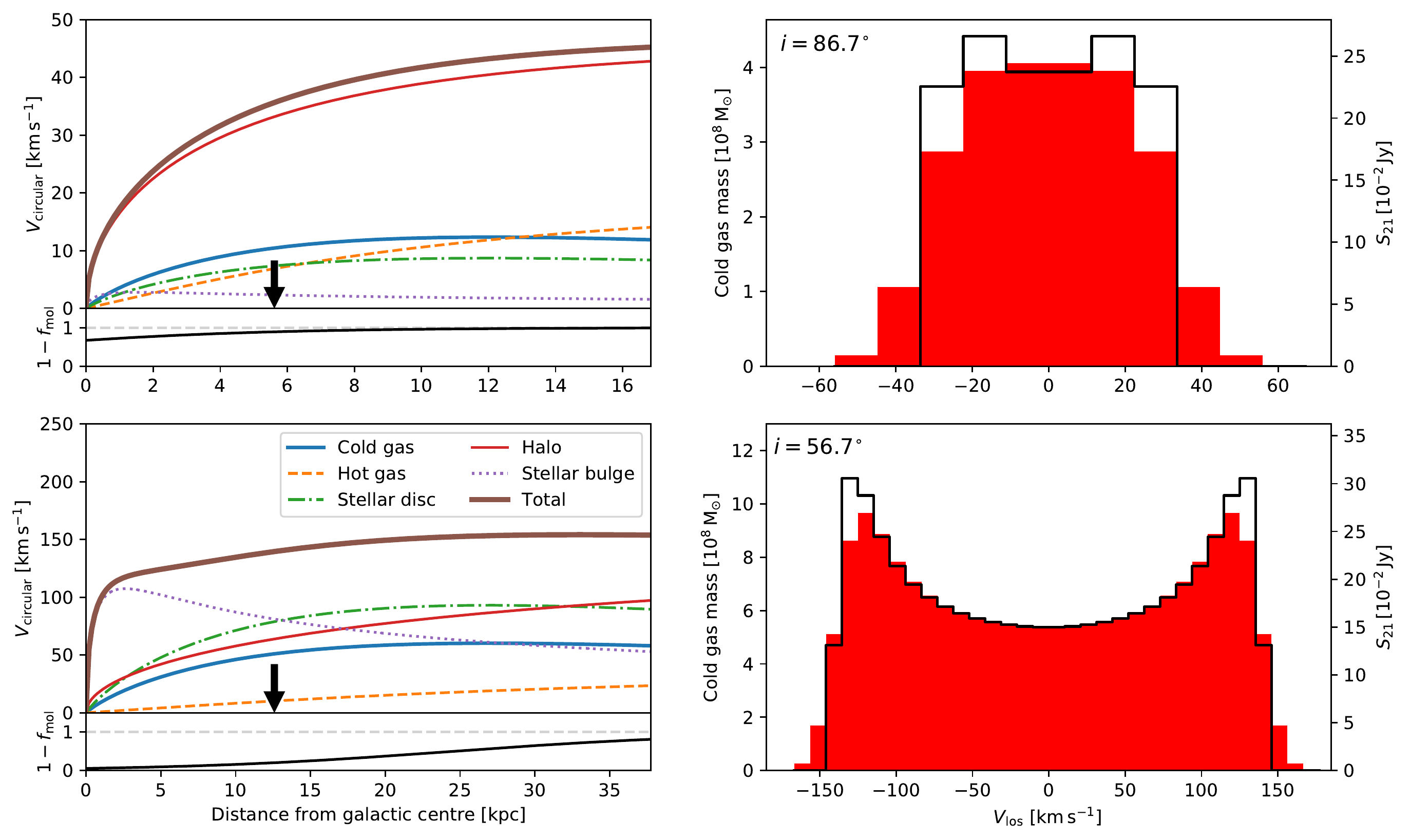}
    \caption{Two examples illustrating our model for generating a mock {21-cm} velocity spectrum for \emph{Sibelius-DARK} + \emph{GALFORM} galaxies. \emph{Left panels:} examples of the circular velocity curve shown within $3\,r_{\mathrm{disc}}$ for a ‘low' line width (top) and an ‘intermediate' line width galaxy (bottom), including the $(1 - f_\mathrm{mol})$ atomic fraction profile beneath. The arrow indicates the scale-length of the cold gas disc, $r_{\mathrm{disc}}$. Contributions from the dark matter halo (red line), stellar bulge (purple dotted line), stellar disc (green dash-dot line), hot gas (orange dashed line) and cold gas (blue solid line) are shown. The sum in quadrature of these components gives the model velocity curve (thick brown line). \emph{Right panels:} the {21-cm} velocity spectrum resulting from the circular velocity curves and H\,\textsc{i} gas surface density profiles for both the case before (black open histogram) and after (red filled histogram) including the gas velocity dispersion, $\sigma_v$. The left-hand y-axis shows the distribution by H\,\textsc{i} mass, while the right-hand y-axis shows the distribution by {21-cm} flux, $S_{21}$. The randomly chosen inclination to the line of sight $i$, for each galaxy is given in the top-left corners of the panels. ‘Low' line width galaxy properties: $\mathrm{M_{vir}} = 4.4\times10^{10}\,\mathrm{M_{\odot}}$, stellar disc mass, $\mathrm{M_{\star, disc}} = 1.3\times10^{8}\,\mathrm{M_{\odot}}$, stellar bulge mass, $\mathrm{M_{\star, bulge}} = 1.2\times10^{7}\,\mathrm{M_{\odot}}$, hot gas mass, $\mathrm{M_{hot}} = 1.1\times10^{10}\,\mathrm{M_{\odot}}$ and cold gas mass, $\mathrm{M_{cold}} = 6.9\times10^{8}\, \mathrm{M_{\odot}}$. ‘Intermediate' line width galaxy properties: $\mathrm{M_{vir}} = 1.4\times10^{12}\,\mathrm{M_{\odot}}$, $\mathrm{M_{\star, disc}} = 3.3\times10^{10}\,\mathrm{M_{\odot}}$, $\mathrm{M_{\star, bulge}} = 2.8\times10^{10}\,\mathrm{M_{\odot}}$, $\mathrm{M_{hot}} = 1.3\times10^{11}\,\mathrm{M_{\odot}}$, $\mathrm{M_{cold}} = 3.7\times10^{10}\, \mathrm{M_{\odot}}$.}
    \label{fig:rotncurve_spectra_examples_sigmaVcorrect}
\end{figure*}

\begin{figure*}
    \centering
	\includegraphics[width=\linewidth]{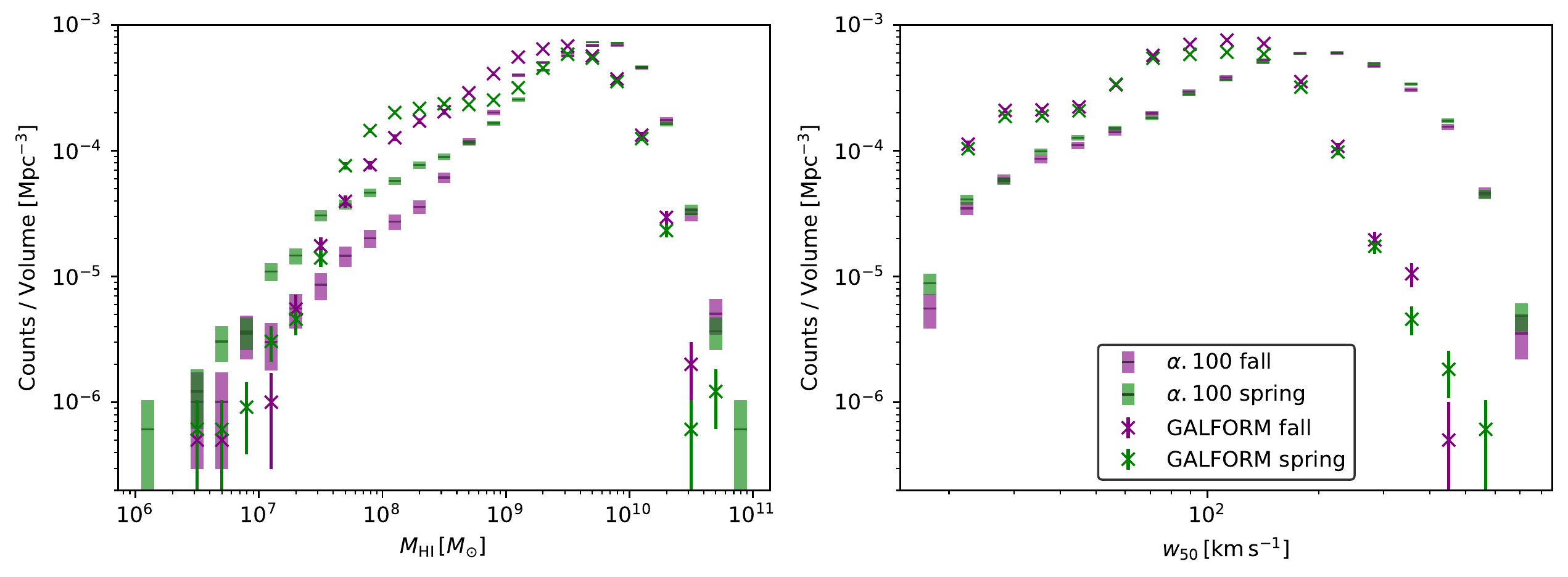}
    \caption{\emph{Left panel:} the observed counts per unit volume as a function of H\,{\textsc{i}} mass for the $\alpha.100$ (dashes) and \emph{GALFORM} (crosses) catalogues. The spring (green) and fall (purple) fields are shown separately. Shaded bands and error bars show Poisson uncertainties. \emph{Right panel:} similar to the left panel but showing the observed counts per unit volume as a function of {21-cm} line width, $w_{50}$.}
    \label{fig:obs_HIMF_HIWF}
\end{figure*}

Four steps are required to construct a mock {21-cm} survey from the \emph{Sibelius-DARK} + \emph{GALFORM} simulation: calculation of the galactic circular velocity curve, determination of the amount of H\,{\textsc{i}} gas as a function of line-of-sight velocity to produce a {21-cm} spectrum, convolution with a kernel to model the thermal broadening of the {21-cm} line, and application of the selection criteria consistent with the chosen {21-cm} survey.

To define the radial mass profiles of galaxies, we follow \citet{2016MNRAS.462.3854L}. The relevant mass-bearing components of a galaxy are the dark matter halo, stellar bulge, stellar disc, hot gas and cold gas. Dark matter haloes are assumed to follow an NFW profile \citep{1996ApJ...462..563N,1997ApJ...490..493N} described by a virial\footnote{We define virial quantities by a sphere enclosing an overdensity that is 200 times the critical density of the Universe, $\rho_{\mathrm{crit}} = 3H^{2}/8\pi G$, and denote them with a ‘vir’ subscript.} mass and concentration. Stellar bulges follow a ‘de~Vaucouleurs' ($r^{1/4}$) law for their surface mass density profile. This profile would require numerical integration to obtain an enclosed mass profile. We follow \citet{2016MNRAS.462.3854L} and assume the simpler, though very similar in shape, \citet{1990ApJ...356..359H} mass profile for bulges instead. The stellar disc is modelled as an infinitesimally thin disc with an exponentially decaying density profile with a half-mass radius $r_{\mathrm{disc}}$. The hot gas is assumed to settle into a spherically symmetric distribution with density profile:
\begin{equation}
    \rho_\mathrm{hot}(r) \propto \frac{1}{(r^2 + r_{c}^{2})^{2}}
\end{equation}
with gas core radius $r_c = 0.1\,r_\mathrm{vir}$, where $r_\mathrm{vir}$ is the virial radius of the dark matter halo. The cold gas is modelled as an infinitesimally thin disc with exponentially decaying surface density profile with the same half-mass radius as the stellar disc, $r_{\mathrm{disc}}$. This is partitioned into atomic and molecular gas components, expressed as the fraction of cold gas that is molecular $f_{\mathrm{mol}} = \Sigma_{\mathrm{mol}}/(\Sigma_{\mathrm{mol}} + \Sigma_{\mathrm{atom}})$, calculated assuming vertical hydrostatic pressure equilibrium and a gas velocity dispersion of $\sigma_{\mathrm{gas}} = 10 \pm 2 \,\mathrm{km\,s^{-1}}$ \citep[for a detailed description, see][]{2011MNRAS.418.1649L}. The atomic hydrogen surface density is assumed to be  $X_\mathrm{H} = 0.74$ times the atomic gas surface density. The galaxy components are assumed to be azimuthally symmetric. Circular velocities for spherically symmetric components are calculated as $v_{c}^{2}(r) = GM(<r)/r$, while for thin exponential discs we use: 
\begin{equation}
v_{c}^{2}=\frac{2GM_\mathrm{disc}}{r_\mathrm{disc}}y^2\left[I_0(y)K_0(y)-I_1(y)K_1(y)\right]
\end{equation}
where $M_\mathrm{disc}$ is the total mass of the disc, $y=r/2r_\mathrm{disc}$, and $I_\nu$ and $K_\nu$ are the modified Bessel functions. The total circular velocity at any radius is then the sum in quadrature over all components. The resulting circular velocity curve, with its decomposition into different components, is shown in the left panels of Fig.~\ref{fig:rotncurve_spectra_examples_sigmaVcorrect} for two example galaxies.

\begin{figure*}
    \centering
	\includegraphics[width=0.8\textwidth]{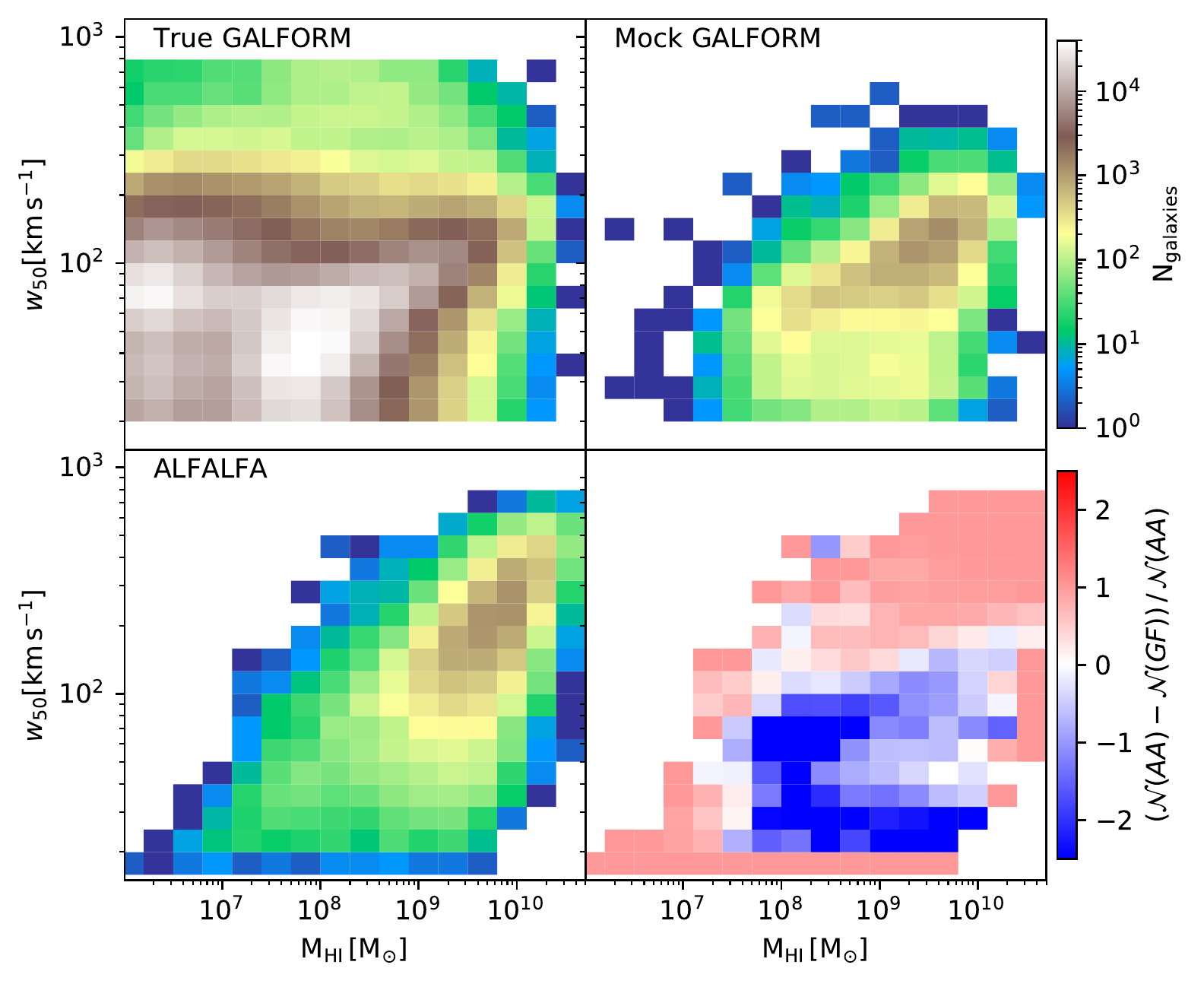}
    \caption{\emph{Upper left:} Number of galaxies across the $w_{50}-M_{\mathrm{HI}}$ plane in the \emph{Sibelius-DARK} + \emph{GALFORM} catalogue, including all galaxies (detected and undetected) in the simulation at $z=0$. The two peaks in the distribution correspond to the contributions from central (peak near $(M_{\mathrm{HI}},w_{50})\sim(10^{8}\,\mathrm{M}_\odot,\,50\,\mathrm{km}\,\mathrm{s}^{-1})$) and satellite galaxies (peak near $(M_{\mathrm{HI}},w_{50})\sim(2\times10^{6}\,\mathrm{M}_\odot,\,70\,\mathrm{km}\,\mathrm{s}^{-1})$). The colour scale represents the number of galaxies per bin in $w_{50}-M_{\mathrm{HI}}$. \emph{Upper right:} Number of galaxies across the $w_{50}-M_{\mathrm{HI}}$ plane in the mock \emph{GALFORM} catalogue (sources passing the ALFALFA selection criteria). The colour scale is the same as in the upper left panel. \emph{Lower left:} Number of galaxies across the $w_{50}-M_{\mathrm{HI}}$ plane in the ALFALFA catalogue.  The colour-scale is the same as in the upper left and upper right panels. \textit{Lower right}: Relative abundance of sources between the ALFALFA and mock \emph{GALFORM} surveys across the $M_\mathrm{HI}-w_{50}$ plane. The two distributions are normalised by their respective integrals across the entire $w_{50}-M_{\mathrm{HI}}$ plane, subtracted in each bin, then compared to the normalised abundance of ALFALFA sources in the same bin. Red regions indicate an excess of ALFALFA sources; blue indicates an excess in \emph{GALFORM}.}
    \label{fig:obs_2Dmhiw50}
\end{figure*}

We define the {21-cm} velocity spectrum as the H\,\textsc{i} mass-weighted distribution of line-of-sight velocities. To determine the line-of-sight velocities, the inclination $i$ of the galaxy to the observer must be set. This is done by drawing a uniformly distributed random value for $\cos(i)$ between $0$ and $1$. We have repeated our analysis using $10$ unique random seeds to assign galaxy inclinations. These cause only small differences of about $8$~per~cent or less in the number of galaxies detected in our mock survey at fixed $w_{50}$.

To calculate how much H\,\textsc{i} gas contributes at each line-of-sight velocity, the model H\,\textsc{i} disc is discretized in the radial (50 bins from $r = 0$ to $r = 4\,r_\mathrm{{disc}}$) and azimuthal angular (45 bins from $\phi = 0$ to $\phi = 2\pi$) coordinates. The line-of-sight velocity of each radial-angular element is $v_{\mathrm{circular}}(r)\sin i \cos \phi$. The radial extent is sufficient to enclose $\geq 90$~per~cent of the cold gas mass, enough to obtain a converged value for $w_{50}$. We always calculate the total flux of galaxies using the total atomic hydrogen mass, integrated to infinity. We match the effective ALFALFA spectral resolution of $10\,\mathrm{km\,s^{-1}}$.

We also account for the velocity dispersion of the H\,\textsc{i} gas. The {21-cm} line is thermally broadened, hence influencing the measured value of the {21-cm} line width. Following \citet{2011MNRAS.418.1649L}, we assume an empirically determined amplitude for the H\,\textsc{i} velocity dispersion of $\sigma_{v} = 10\,\pm 2\,\mathrm{km\,s^{-1}}$. \emph{GALFORM} also assumes the gas velocity dispersion to be $10\,\pm\,2\,\mathrm{km\,s^{-1}}$ \citep[see,][and references therein]{2016MNRAS.462.3854L}. We implement this by convolving the spectrum with a Gaussian kernel of standard deviation, $\sigma_v$. The right-hand panels in Fig.~\ref{fig:rotncurve_spectra_examples_sigmaVcorrect} show {21-cm} velocity spectra for the same two example galaxies as shown in the left panels. The example in the upper panels is a galaxy where the velocity dispersion is comparable to the maximum circular velocity. The example in the lower panels is a galaxy where the velocity dispersion is much smaller than the maximum circular velocity. In both cases, the black curves show the spectra before convolution with the thermal broadening kernel, while the red bins show the spectra including thermal broadening. We measure the {21-cm} line width, $w_{50}$, as the full width at half maximum of the spectra. We choose $w_{50}$ over alternatives such as the full width at 20~per~cent maximum, $w_{20}$, because $w_{50}$ is less sensitive to noise in the {21-cm} spectra, enabling more sources to be used \citep{2010MNRAS.403.1969Z}.

To replicate the ALFALFA survey in order to study the observational and statistical effects on the measurements of the H\,{\textsc{i}}\,WF, we first need to create a mock catalogues of \emph{Sibelius-DARK} + \emph{GALFORM} sources that would have been detected by an ALFALFA-like survey. We draw our initial set of galaxies from the $z=0$ output of the \emph{GALFORM} model.\footnote{The ALFALFA survey detects galaxies out to $z\sim0.05$. We have repeated our analysis using the $z=0.05$ output of the same \emph{GALFORM} model and find only small differences of about $5$~per~cent or less in the number of galaxies detected in our mock survey at fixed $w_{50}$.}

We apply two criteria to determine which galaxies are included in our mock survey catalogue. First, only sources in the ALFALFA survey footprints should be included \citep[][tables D1-D4]{2018MNRAS.477....2J}. The \emph{Sibelius-DARK} halo catalogues provide (RA, Dec) coordinates which we use to determine whether a source is included within the footprints. Second, the ALFALFA CL must be applied. We implement a continuous CL by linearly interpolating between the $25$, $50$, and $90$~per~cent spring and fall CLs given in equations~(\ref{90_CL_S}-\ref{25_CL_F}), and linearly extrapolating to $0$ and $100$~per~cent completeness. We draw a uniformly distributed random number between $0$ and $1$ for each \emph{GALFORM} galaxy. If this number exceeds the survey completeness for the given {21-cm} line-width and flux of the galaxy, the galaxy is discarded from the catalogue. The result is a mock \emph{GALFORM} survey containing $21,031$ sources, of which $12,408$ are in the spring field and $8,623$ are in the fall field (see Table~\ref{table:galaxy_characteristics}, which additionally includes the total number of \emph{GALFORM} sources before the selection criteria are applied). The distribution of the mock survey sources in (RA, Dec) and distance is overlaid on the ALFALFA survey source distributions in Fig.~\ref{fig:footprints_coneplots_subplots}.

For a survey like ALFALFA, there can be instances whereby two
or more galaxies fall inside the beam at the same time and overlap
in frequency, an effect termed ‘beam confusion’. Our method to construct a mock survey does not account for beam confusion. This could plausibly be a limitation when comparing our mock survey to observations. \citet{2013ApJ...766..137O} found that ‘confused' galaxies typically have high H\,{\textsc{i}} mass and $w_{50}$, with $M_{\mathrm{HI}} > 10^{10}\,\mathrm{M}_\odot$ and $w_{50} > 300\,\mathrm{km}\,\mathrm{s}^{-1}$, albeit for the HIPASS
survey, which has a larger beam than ALFALFA. Subsequently, \citet{2019MNRAS.488.5898C} found lower levels of confusion in these ranges for their mock ALFALFA surveys and that any confusion only reduced the total number of galaxies in their sample by less than $1$~per~cent. \citet{2015MNRAS.449.1856J} also found that beam confusion can only slightly change the shape of the H\,\textsc{i}\,MF, by no more than would already be allowed by the random errors on the measurements. It therefore seems unlikely that beam confusion is one of the main drivers of the systematic difference between the spring and fall H\,\textsc{i}\,WFs, so we omit further discussion of this effect from our analysis below.

Fig.~\ref{fig:obs_HIMF_HIWF} shows the counts in the catalogues per unit volume as a function of H\,{\textsc{i}} mass (left panel) and $w_{50}$ (right panel) for the ALFALFA $\alpha.100$ catalogue (dashes) and \emph{Sibelius-DARK} + \emph{GALFORM} (crosses) for the spring (green) and fall (purple) fields individually. The \emph{GALFORM} H\,{\textsc{i}} mass distribution overall appears similar in shape to that of ALFALFA. Noticeable differences include the overdensity of \emph{GALFORM} sources at intermediate masses, $10^8\lesssim M_{\mathrm{HI}}/\mathrm{M}_\odot \lesssim 10^9$, and the underdensity of sources at the highest H\,{\textsc{i}} masses. The \emph{GALFORM} $w_{50}$ distribution, on the other hand, has a starkly different shape to that measured in ALFALFA. There is an overdensity of sources for line widths below $\sim150\,\mathrm{km}\,\mathrm{s}^{-1}$, and an underdensity above. We comment further on these differences in Sec.~\ref{sec:Discussion}.

The distribution of sources in the $w_{50}-M_{\mathrm{HI}}$ plane for the ‘true \emph{GALFORM}' data (detected and undetected galaxies in the simulation), the ‘mock \emph{GALFORM}' data (only those simulation sources within the ALFALFA selection criteria) and the ALFALFA data are shown in Fig.~\ref{fig:obs_2Dmhiw50}. The true \emph{GALFORM} data displays a bi-modal distribution of sources which can be attributed to the relative contribution from satellite and central galaxies, respectively. There are obvious differences in the distribution between the true \emph{GALFORM} and the mock, or real, ALFALFA survey. Given that the simulation will produce many galaxies that are either low in flux, and/or possess large values for $w_{50}$, these sources will not satisfy the ALFALFA selection criteria and hence ‘drop out' of the final survey catalogue. Alongside these $w_{50}-M_{\mathrm{HI}}$ distributions, we show the relative abundance of sources between the ALFALFA and mock \emph{GALFORM} surveys in the bottom-right panel of Fig.~\ref{fig:obs_2Dmhiw50}. Generally, the ALFALFA survey contains more sources at the edges of the $w_{50}-M_{\mathrm{HI}}$ distribution. In particular, the smallest bin in $w_{50}$ ($w_{50} \sim\,20-30 \mathrm{km}\,\mathrm{s}^{-1}$) contains no sources for our \emph{GALFORM} mock. This comes from a limitation in our method when accounting for thermal broadening of the 21-cm emission line; see Sec~\ref{sec:mock_survey} for details. On the other hand, the mock \emph{GALFORM} survey has an abundance of sources relative to ALFALFA for $(M_{\mathrm{HI}},w_{50})\sim(10^{8}-10^{10}\,\mathrm{M}_\odot,\,25-150\,\mathrm{km}\,\mathrm{s}^{-1})$. This is consistent with the corresponding 1D $M_{\mathrm{HI}}$ and $w_{50}$ distributions shown in Fig.~\ref{fig:obs_HIMF_HIWF}. We comment further on differences in the $w_{50}-M_{\mathrm{HI}}$ plane between our mock and the real ALFALFA survey in Sec.~\ref{sec:mHIw50_distribution}. 

\begin{figure*}
    \centering
	\includegraphics[width=\textwidth]{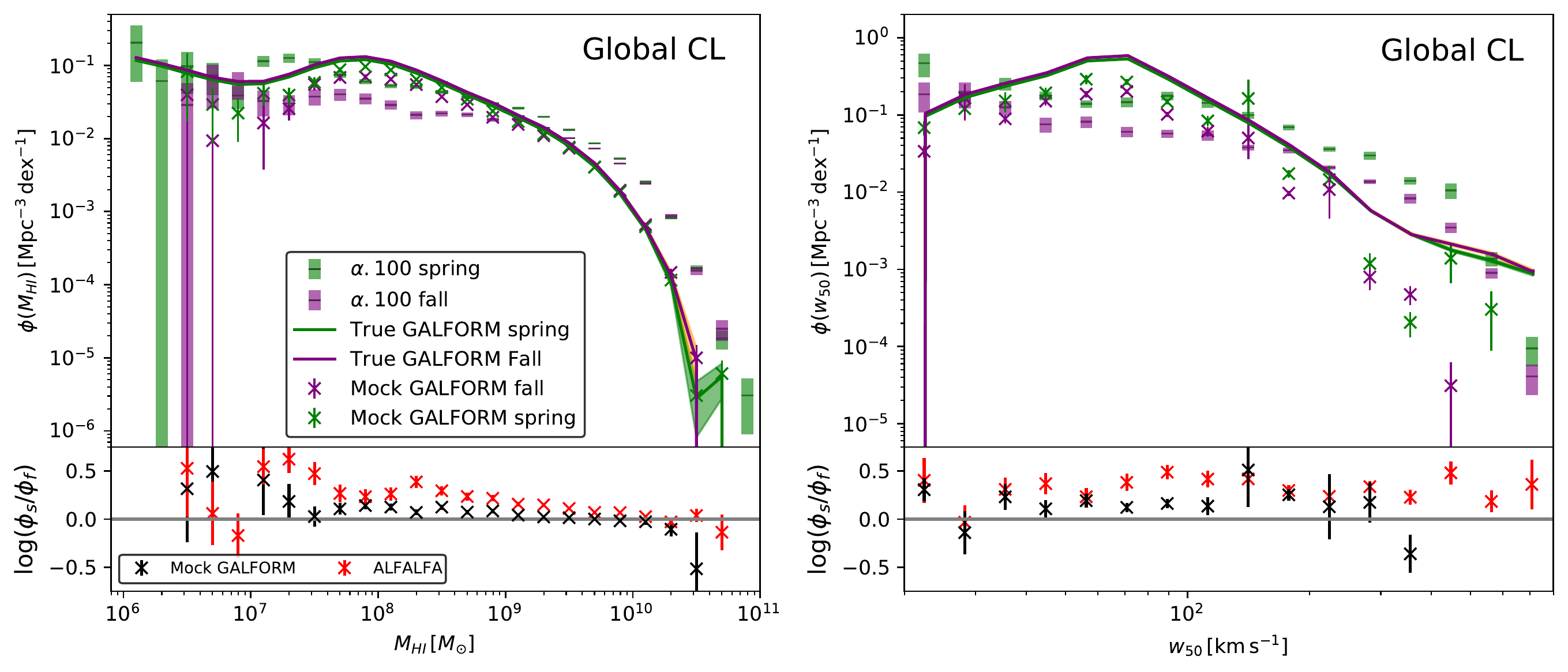}
    \includegraphics[width=\textwidth]{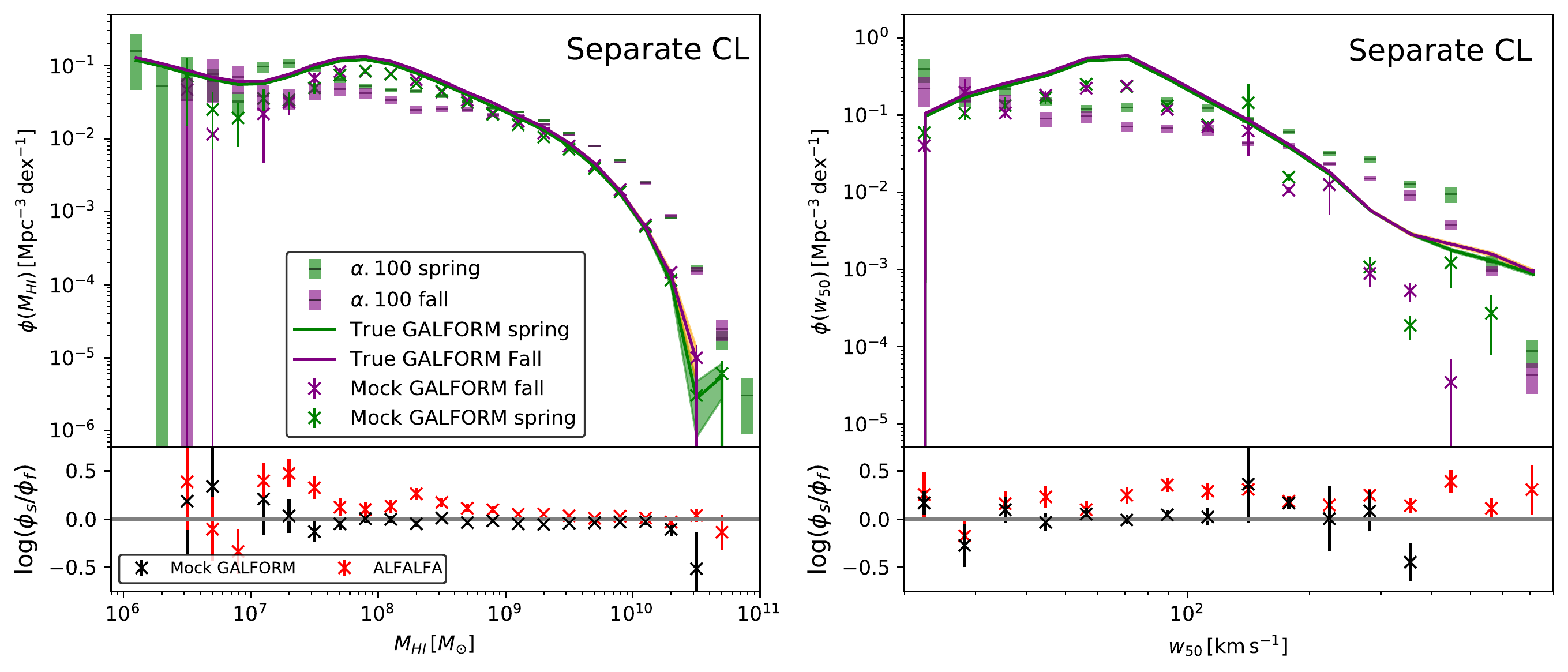}
    \caption{\emph{Upper left panel:} the H\,\textsc{i} mass function (H\,\textsc{i}\,MF) calculated by adopting the same completeness limit in the spring and fall fields. The main panel shows the H\,\textsc{i}\,MF measured from the $\alpha.100$ (dashes with shaded 1$\sigma$ uncertainties) and \emph{GALFORM} catalogues (crosses with 1$\sigma$ uncertainties shown with error bars), separately for the two ALFALFA fields, spring (green) and fall (purple). Additionally, the ‘true' \emph{GALFORM} H\,\textsc{i}\,MF (solid lines with 1$\sigma$ uncertainty shown with shaded band) is shown for the spring and fall fields. The lower sub-panel shows the ratio of the spring and fall \emph{GALFORM} (black crosses) and ALFALFA (red crosses) H\,\textsc{i}\,MFs with 1$\sigma$ uncertainties. \emph{Upper right panel:}  The H\,\textsc{i} width function (H\,\textsc{i}\,WF) calculated by adopting the same completeness limit in the spring and fall fields. Lines and symbols are as in upper left panel. \emph{Lower left and right panels:} similar to upper panels, but showing the H\,\textsc{i}\,MF and H\,\textsc{i}\,WF calculated by adopting separate completeness limits in the spring and fall fields.}
    \label{fig:HIMF_HIWF}
\end{figure*}

\subsection{\texorpdfstring{The $1/V_\mathrm{{\lowercase{eff}}}$ maximum likelihood estimator}{The 1/Veff maximum likelihood estimator}}\label{sec:veff_estimator} 

The $1/V_\mathrm{eff}$ estimator is used to estimate the abundance of undetected galaxies with a given $M_\mathrm{HI}$ and $w_{50}$ from the abundance of those galaxies that were detected. Galaxies can be undetected in the survey due to being low-mass, and hence low-flux sources and/or due to having wider line widths which spread their emission over more channels in the detector. The original SWML estimator \citep{1988MNRAS.232..431E} is applicable to galaxy samples that are integrated-flux limited. The SWML can be extended to become the bivariate stepwise maximum likelihood (2DSWML) estimator for surveys that have a selection function that depends on two observable quantities, such as ALFALFA. The only difference between the $1/V_\mathrm{{eff}}$ and 2DSWML estimators is that in the $1/V_\mathrm{eff}$ case, the effective volumes are iteratively calculated for each individual galaxy, instead of calculated per 2D bin in $M_\mathrm{{HI}}$-$w_{50}$ space as in the 2DSWML approach \citep{2005MNRAS.359L..30Z,2010ApJ...723.1359M}. The effective volumes found for each galaxy are maximum likelihood counterparts of the classical $1/V_\mathrm{max}$ volumes \citep{1968ApJ...151..393S} with the important difference that the $1/V_\mathrm{eff}$ method in principle corrects for spatial non-uniformity in the source distribution.

The $1/V_\mathrm{eff}$ estimator is 2D in the sense that the CL of the ALFALFA survey depends both on the integrated {21-cm} flux, $S_{21}$ and line-width, $w_{50}$, of the source. The implementation of the $1/V_\mathrm{eff}$ estimator is the same as in \cite{2022MNRAS.509.3268O} with the only difference being that we adopt separate CLs for the two survey fields. The $1/V_\mathrm{eff}$ method requires knowledge of the survey CL in order to produce the 2D H\,\textsc{i}\, mass-width function. Instead of using the global $50$~per~cent CL as in \citet[][equation~A5]{2022MNRAS.509.3268O}, we adopt the $50$~per~cent CL appropriate to each survey field (Equations~\ref{50_CL_S} \& \ref{50_CL_F}). From the summation of the values of the $1/V_\mathrm{eff}$ weights in 2D bins in $M_{\mathrm{HI}}$ and $w_{50}$ we compute the 2D H\,\textsc{i}\, mass-width function. The sum along the $w_{50}$ axis gives the H\,\textsc{i}\,MF, while the same along the mass axis gives the H\,\textsc{i}\,WF.

\section{\texorpdfstring{ALFALFA and \emph{Sibelius-DARK} + \emph{GALFORM} H\,\textsc{i} width functions}{ALFALFA and Sibelius-DARK + GALFORM HI width functions}}\label{sec:results_HIWF} 

Understanding the origins of spatial variations of the normalisation of the ALFALFA H\,\textsc{i}\,WF is crucially important in the context of using it as a constraint on cosmology. The $\Lambda$CDM cosmological model predicts that the dark matter HMF should be universal \citep[in shape and normalisation, e.g.][]{1988ApJ...327..507F, 1996MNRAS.282..347M, 2002MNRAS.329...61S, 2009MNRAS.399.1773C} and therefore similar in the two fields surveyed in ALFALFA, because the volumes sampled are sufficiently large. We have checked this explicitly in the \emph{Sibelius-DARK} simulation: the HMFs in the spring and fall volumes differ by no more than $8$~per~cent (within their uncertainties) at any halo mass $10^{8}<M_{\mathrm{vir}}/\mathrm{M}_\odot<10^{14}$ (see Sec.~\ref{sec:sibelius}). 

The most straightforward prediction for the H\,\textsc{i}\,WF is that it should also have the same shape and normalisation (within about 8~per~cent) in the two fields \citep[see Sec.~5.3.2,][for a detailed account of the connection between the HMF and the H\,\textsc{i}\,WF]{2022MNRAS.509.3268O}. Indeed, the \emph{Sibelius-DARK} + \emph{GALFORM} galaxy catalogue confirms this. In the right panels of Fig.~\ref{fig:HIMF_HIWF} we show the H\,\textsc{i}\,WF of all galaxies with $M_{\mathrm{HI}}>10^{6}\,\mathrm{M}_{\odot}$ in the spring and fall survey fields (regardless of whether they would be detected) with the green and purple solid lines, respectively. Analogously to the HMFs in the two regions, the two H\,\textsc{i}\,WF curves differ by no more than $8$~per~cent at any line width. If the approximately factor of $2$ difference in normalisation of the H\,\textsc{i}\,WF for the spring and fall ALFALFA fields cannot otherwise be explained, then the $\Lambda$CDM model could be called into question \citep[][where the first entry highlights the issue in the HIPASS survey and those after for ALFALFA]{2010MNRAS.403.1969Z, 2011ApJ...739...38P, 2022MNRAS.509.3268O}. We have already outlined above (Sec.~\ref{sec:mock_survey}) that random scatter in the HIWF driven by random inclination angles and the systematic effects due to the redshift evolution of galaxies within the survey volume are much too small to explain the large observed difference in normalisation, so we now turn our attention to other potential sources of error.

In Sec.~\ref{sec:Global_CL} we measure the H\,{\textsc{i}}\,WF of our mock \emph{Sibelius-DARK} + \emph{GALFORM} survey using exactly the same method as \citet{2022MNRAS.509.3268O}, and compare to their measurement for the ALFALFA survey. Next, in Sec.~\ref{sec:Separate_CL}, we repeat the measurement assuming the separately derived CLs for the spring and fall fields in the calculation of the $1/V_\mathrm{eff}$ weights, and compare to the same approach applied to the $\alpha.100$ catalogue. In Sec.~\ref{sec:LoSClustering} we investigate effect of the clustering of sources along the line of sight upon the H\,{\textsc{i}}\,WFs. Finally, in Sec.~\ref{sec:mHIw50_distribution} we investigate the effect of the galaxy distribution in the H\,\textsc{i} mass -- spectral line width plane.

\subsection{Fiducial analysis of the mock surveys}\label{sec:Global_CL}

We make the measurement of the H\,\textsc{i}\,MF and H\,\textsc{i}\,WF for the spring and fall mock \emph{Sibelius-DARK} + \emph{GALFORM} catalogues separately, following the procedure outlined in \citet{2022MNRAS.509.3268O} where the globally derived CL for the ALFALFA survey \citep[][equation~A5]{2022MNRAS.509.3268O} is assumed in the $1/V_\mathrm{{eff}}$ estimator. The top-left and right panels of Fig.~\ref{fig:HIMF_HIWF} respectively show the H\,\textsc{i}\,MFs and H\,\textsc{i}\,WFs measured using this approach. The measurements of \citet{2022MNRAS.509.3268O}, now adapted to retain only sources within $200\,\mathrm{Mpc}$ (Sec.~\ref{sec:ALFALFA_Survey}), are also shown in these panels for comparison.

We restrict ourselves to a qualitative comparison of the $\alpha.100$ ALFALFA (dash marker with shaded box for the spring and fall fields in green and purple, respectively) and ‘mock \emph{GALFORM}' (crosses of corresponding colours) H\,{\textsc{i}}\,WFs. We find that the ALFALFA and mock \emph{GALFORM} H\,{\textsc{i}}\,WFs have similar, almost constant low line-width slopes for $w_{50} \lesssim 100\,\mathrm{km}\,\mathrm{s}^{-1}$. For larger $w_{50}$, \emph{GALFORM} underpredicts the number density of sources significantly.

Despite the ‘true \emph{GALFORM}' curves for the spring and fall fields (green and purple solid lines) differing by less than $8$~per~cent across the entire range in $w_{50}$, we find that our mock {21-cm} survey has a qualitatively similar offset between the spring and fall H\,\textsc{i}\,WFs as is observed in ALFALFA. The number density in the spring field exceeds that in the fall field throughout the line-width range. The median ratio between the spring to fall H\,\textsc{i}\,WFs is $\log_{10}(\phi_{s}\,/\,\phi_{f}) = 0.36 \pm 0.09$ ($16^{\mathrm{th}}$--$84^{\mathrm{th}}$ percentile scatter about the median) for ALFALFA, and $0.17 \pm 0.13$ for mock \emph{GALFORM}. Our mock \emph{GALFORM} H\,\textsc{i}\,WFs capture the shapes of their true counterparts reasonably well, although they do underestimate them by up to a factor of 3 at $w_{50} \lesssim 100\,\mathrm{km}\,\mathrm{s}^{-1}$, and up to a factor of 10 at higher line widths. The reason for this is that there are combinations of \ion{H}{i} masses and line widths where galaxies exist, but none are observed (because the survey is not sensitive to detect them); this issue is discussed in detail by \citet[][sec.~5.2.6]{2022MNRAS.509.3268O}.

We focus our attention on the puzzling result that the mock \emph{GALFORM} H\,\textsc{i}\,WFs are systematically offset from each other in the same sense as ALFALFA. We first consider the influence of the choice of CL for each survey field.

\subsection{Influence of the survey completeness limit}\label{sec:Separate_CL}

We repeat the measurement of the H\,\textsc{i}\,MF and H\,\textsc{i}\,WF for the spring and fall $\alpha.100$ and mock \emph{Sibelius-DARK} + \emph{GALFORM} catalogues as in Sec.~\ref{sec:Global_CL}, but this time assuming CLs derived separately for the spring and fall ALFALFA fields (equations~\ref{50_CL_S} \& \ref{50_CL_F}) in the calculation of the $1/V_{\mathrm{eff}}$ weights. The bottom-right panel of Fig.~\ref{fig:HIMF_HIWF} shows the measured H\,\textsc{i}\,WF using this approach.

The $\alpha.100$ ALFALFA and mock \emph{GALFORM} H\,\textsc{i}\,WFs have qualitatively similar shapes to those discussed in Sec.~\ref{sec:Global_CL}. However,  the spring and fall mock survey H\,\textsc{i}\,WFs are no longer systematically offset from each other. Therefore, when the same CLs used to construct the mock catalogues (Sec.~\ref{sec:veff_estimator}) are used to correct them for incompleteness, the $1/V_\mathrm{eff}$ algorithm correctly recovers the fact that the H\,\textsc{i}\,WFs in the two fields have indistinguishable shapes and amplitudes (although the true number densities are still underestimated). Quantifying this, the median ratio between the spring to fall H\,\textsc{i}\,WFs is $\log_{10}(\phi_{s}\,/\,\phi_{f}) = 0.05 \pm 0.14$.

For ALFALFA, the systematic offset between the spring and fall H\,\textsc{i}\,WFs is reduced when separate spring and fall CLs are used in the calculation of the $1/V_\mathrm{eff}$ weights, but does not disappear: the median ratio between the spring to fall H\,\textsc{i}\,WFs is $\log_{10}(\phi_{s}\,/\,\phi_{f}) = 0.25 \pm 0.09$. Because there is empirical evidence that the ALFALFA survey has different CLs in the spring and fall fields \citep{2011AJ....142..170H,2022MNRAS.509.3268O}, our view is that these measurements should be taken to supersede those of \citet{2022MNRAS.509.3268O} when taking the same cut in recessional velocity -- we tabulate them in Appendix~\ref{appendix:tables} for both the H\,\textsc{i}\,MFs and H\,\textsc{i}\,WFs in the spring and fall regions. We also include a combined measurement for the entire survey in which the $1/V_{\mathrm{eff}}$ weights have been derived separately for galaxies in the two regions accounting for the different survey sensitivity in each.

The fact that the spring and fall ALFALFA H\,\textsc{i}\,WFs are still offset from each other after correcting for the different CLs in the two fields suggests that this may not be the only systematic effect influencing the measurement. There are a limited number of other factors that can influence the outcome of the $1/V_{\mathrm{eff}}$ measurement. We have identified only two: differences in the clustering along the line of sight in the two fields and/or differences in the properties of galaxies in the two fields as reflected by their distribution in the 2D space of $M_{\mathrm{HI}}-w_{50}$.

In order to focus on these other effects, for our analysis in the remainder of this section we remove the influence of the CL by imposing a very conservative completeness cut on both the ALFALFA survey and on our mock surveys: we impose a CL offset in $S_{21}$ by $0.10\,\mathrm{dex}$ above that empirically derived for the fall field (equation~\ref{50_CL_F}). We also assume this CL in the calculations of $1/V_{\mathrm{eff}}$ weights below.

\subsection{Influence of the clustering of sources along the line of sight}\label{sec:LoSClustering}

Our use of the $1/V_\mathrm{eff}$ estimator has so far treated the spring and fall fields as fully independent: when calculating the H\,\textsc{i}\,WF and H\,\textsc{i}\,MF for the spring field, no information is available about the large-scale structure in the fall field, and vice versa. We now wish to investigate whether the different clustering along the line of sight in each field causes a spurious systematic offset between the H\,\textsc{i}\,WFs. We proceed as follows. We calculate $1/V_\mathrm{{eff}}$ weights for the full $\alpha.100$ catalogue, thereby obtaining effective volumes for each source without any knowledge of the different clustering of sources along the line of sight in each field. The $1/V_{\mathrm{eff}}$ algorithm makes no use of the $(\mathrm{RA},\mathrm{Dec})$ coordinates of sources, so this approach treats the survey as a single contiguous volume. We then separately compute the H\,\textsc{i}\,WF for each field using the effective volumes for sources in that field. This exercise should erase the offset between the spring and fall fields if the distribution of sources along the line of sight is its direct cause.

\begin{figure}
    \centering
    \includegraphics[width=\columnwidth]{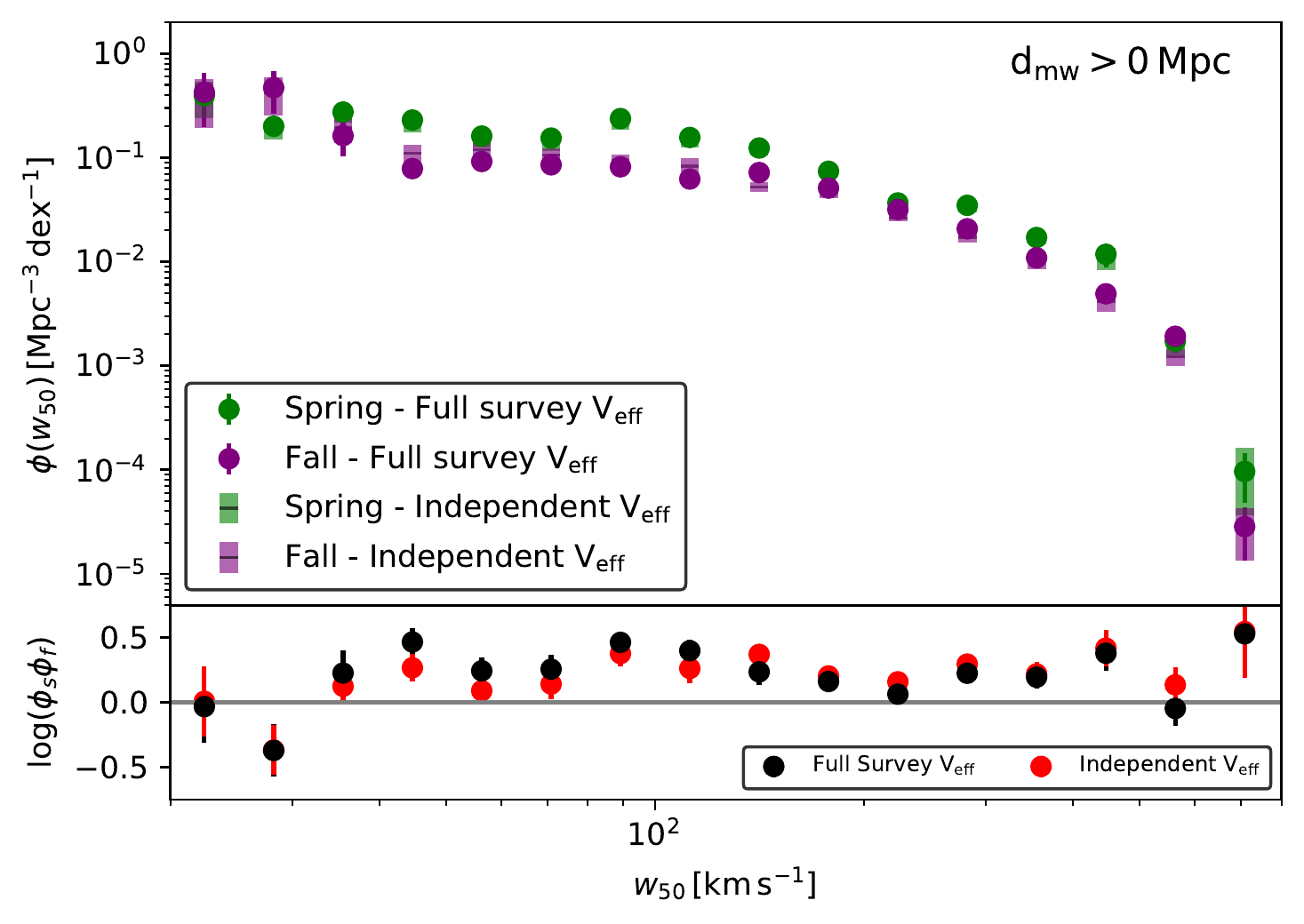}
    \includegraphics[width=\columnwidth]{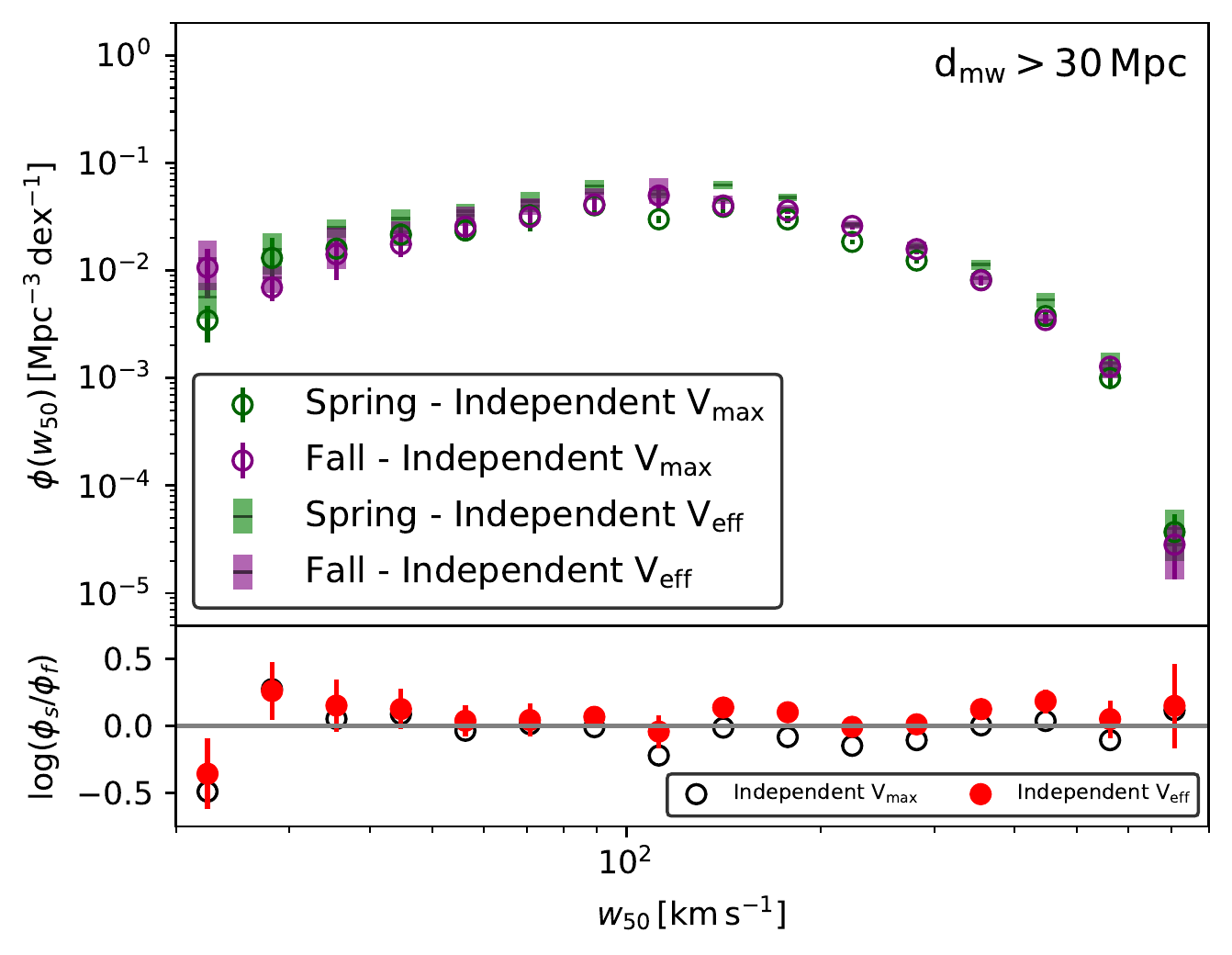}
    \includegraphics[width=\columnwidth]{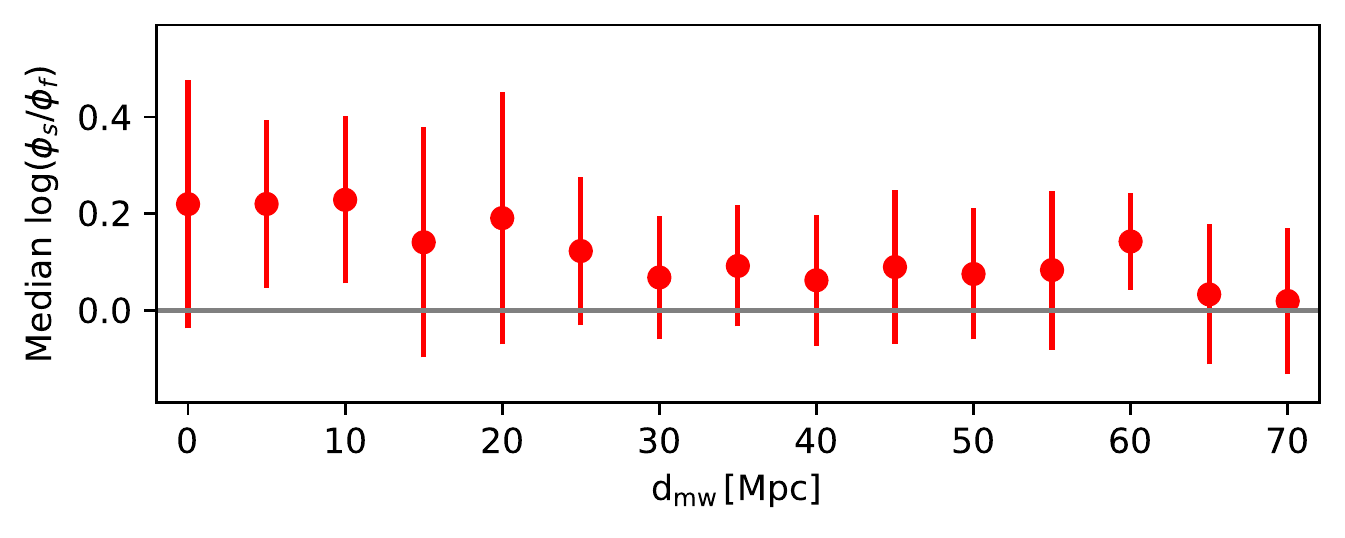}
    \caption{\emph{Upper panel:} The spring (green) and fall (purple) $\alpha.100$ ALFALFA H\,\textsc{i}\,WF as derived using the $1/\mathrm{V_{eff}}$ weights assigned independently (described in Sec.~\ref{sec:LoSClustering}; dashes with shaded 1$\sigma$ uncertainties) and from the full survey catalogue (points with error bars showing 1$\sigma$ uncertainties). The completeness limit adopted has been raised by $0.10\,\mathrm{dex}$ above that empirically derived for the fall field. Sources detected at all distances from the Milky~Way are included. The lower sub panel shows the spring-to-fall ratio for the independent (red points) and full survey (black points) $1/\mathrm{V_{eff}}$ weights. \emph{Middle panel:} The spring (green) and fall (purple) H\,\textsc{i}\,WF  derived using the $1/\mathrm{V_{eff}}$ (dashes) and $1/\mathrm{V_{max}}$ (open circles) weights assigned independently but including only those sources detected at distances $d_{\mathrm{mw}} > 30\,\mathrm{Mpc}$. The lower sub panel again shows the spring-to-fall ratio for the independent $1/\mathrm{V_{eff}}$ (red points) and independent $1/\mathrm{V_{max}}$ (open black circles) $1/\mathrm{V_{eff}}$ weights. \emph{Lower panel:} The median H\,\textsc{i}\,WF ratio for spring-to-fall regions (derived using the $1/\mathrm{V_{eff}}$ weights) as a function of the minimum distance from the Milky Way at which sources are retained, $\mathrm{d_{mw}}$.} 
    \label{fig:AA_HIWF_losclustering}
\end{figure}

The resulting H\,\textsc{i}\,WFs are shown in the top panel of Fig.~\ref{fig:AA_HIWF_losclustering}. We label the case for which the survey fields are treated independently as ‘independent $V_\mathrm{eff}$', and the case for which the survey fields are analysed jointly as ‘full survey $V_\mathrm{eff}$'. We find that there is almost no change in the systematic offset between the spring and fall H\,\textsc{i}\,WFs; the median ratio between the spring to fall H\,\textsc{i}\,WFs derived using the independent $V_\mathrm{eff}$ weights is $\log_{10}(\phi_{s}\,/\,\phi_{f}) = 0.22 \pm 0.11$. Therefore, the different distribution of sources along the line of sight in the two fields is not an important driver of the systematic offset. This leaves the distribution of sources in the 2D $M_{\mathrm{HI}}-w_{50}$ parameter space to be investigated.

\subsection{\texorpdfstring{Influence of the galaxy distribution in $M_{\mathrm{HI}}-w_{50}$}{Influence of the galaxy distribution in MHI-w50}}\label{sec:mHIw50_distribution}

\begin{figure*}
    \centering
	\includegraphics[width=\textwidth]{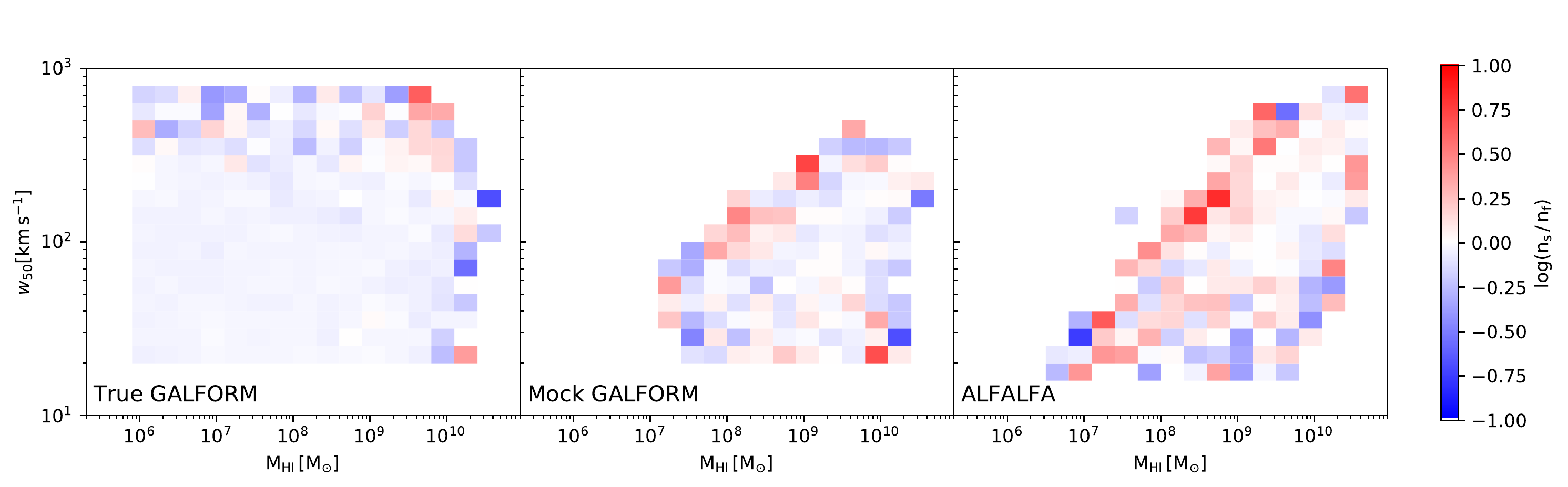}
    \caption{Logarithmic ratio of the abundance per unit volume of sources in the $w_{50}-M_{\mathrm{HI}}$ plane between the spring and fall survey fields. \emph{Left panel:} \emph{Sibelius-DARK} + \emph{GALFORM} catalogue data including all galaxies (detected and undetected). The approximately uniform overdensity in the fall relative to the spring region is due to the different HMFs in the two regions as detailed in Sec.~\ref{sec:sibelius}. \emph{Middle panel:} mock \emph{GALFORM} catalogue data (detected sources only) weighted by $1/V_\mathrm{{eff}}$ derived independently in each survey field. \emph{Right panel:} ALFALFA catalogue data weighted by $1/V_\mathrm{{eff}}$ derived independently in each survey field. In the centre and right panels, only sources above a completeness limit $0.10\,\mathrm{dex}$ above that empirically derived for the fall field are included.}
    \label{fig:AA_GF_w50mHI_sf}
\end{figure*}

The distribution of detected sources in the $M_{\mathrm{HI}}-w_{50}$ plane is the fundamental input into the $1/V_\mathrm{eff}$ estimator. Even if the dark matter HMF is identical in the spring and fall fields, should galaxy formation and evolution proceeded differently in the two fields this could lead to differences between their respective H\,\textsc{i}\,WFs. We therefore check for measurable differences in the $M_{\mathrm{HI}}-w_{50}$ distribution of sources between the two survey fields, which could be a symptom of such differential galaxy evolution. We show ratios of the $M_{\mathrm{HI}}-w_{50}$ distributions in the two fields in Fig.~\ref{fig:AA_GF_w50mHI_sf}. The intrinsic distribution of \emph{GALFORM} galaxies (‘true \emph{GALFORM}', left panel) is slightly overdense in the fall field. The amplitude of the fall overdensity in the $M_{\mathrm{HI}}-w_{50}$ plane is consistent with that in the HMF (see Sec~\ref{sec:sibelius}). We interpret this to imply that galaxy formation in this \emph{GALFORM} model proceeds essentially identically in the two fields. There is also some noise due to low number counts at high $M_{\mathrm{HI}}$ and/or $w_{50}$.

In the centre panel, we show the ratio of the distributions of \emph{GALFORM} galaxies detected in our mock survey, weighted by their $1/V_{\mathrm{eff}}$ weights (calculated separately for the spring and fall fields). Large differences would arise due to the different CLs in the two fields, so we make this comparison after imposing the aforementioned more conservative completeness cut. We repeat the same process for the ALFALFA survey and show the result in the right panel of the figure. In the mock survey, the two distributions agree over most of the parameter space, except around $(M_{\mathrm{HI}},w_{50})\sim(10^{8.5}\,\mathrm{M}_\odot,120\,\mathrm{km}\,\mathrm{s}^{-1})$. The ALFALFA survey has a similar feature at the same location, but also has $\log(n_\mathrm{s}/n_\mathrm{f})>0$ over most of the space. This overall overdensity in the spring field is the same one reflected in the systematically higher number density in the same field in the H\,\textsc{i}\,MF and H\,\textsc{i}\,WF: these functions are the integrals of the H\,\textsc{i} mass-width function along the width and mass axes, respectively. We note that this does not necessarily imply that the bias need be near-uniform across the $M_{\mathrm{HI}}-w_{50}$ plane. Fig.~\ref{fig:AA_GF_w50mHI_sf} therefore suggests that the difference in number density seems to arise across all line widths and H\,\textsc{i} masses and is not tied to any particular region in this parameter space. The local overdensity near $(M_{\mathrm{HI}},w_{50})\sim(10^{8.5}\,\mathrm{M}_\odot,120\,\mathrm{km}\,\mathrm{s}^{-1})$ bears further investigation.

\begin{figure}
    \centering
	\includegraphics[width=\columnwidth]{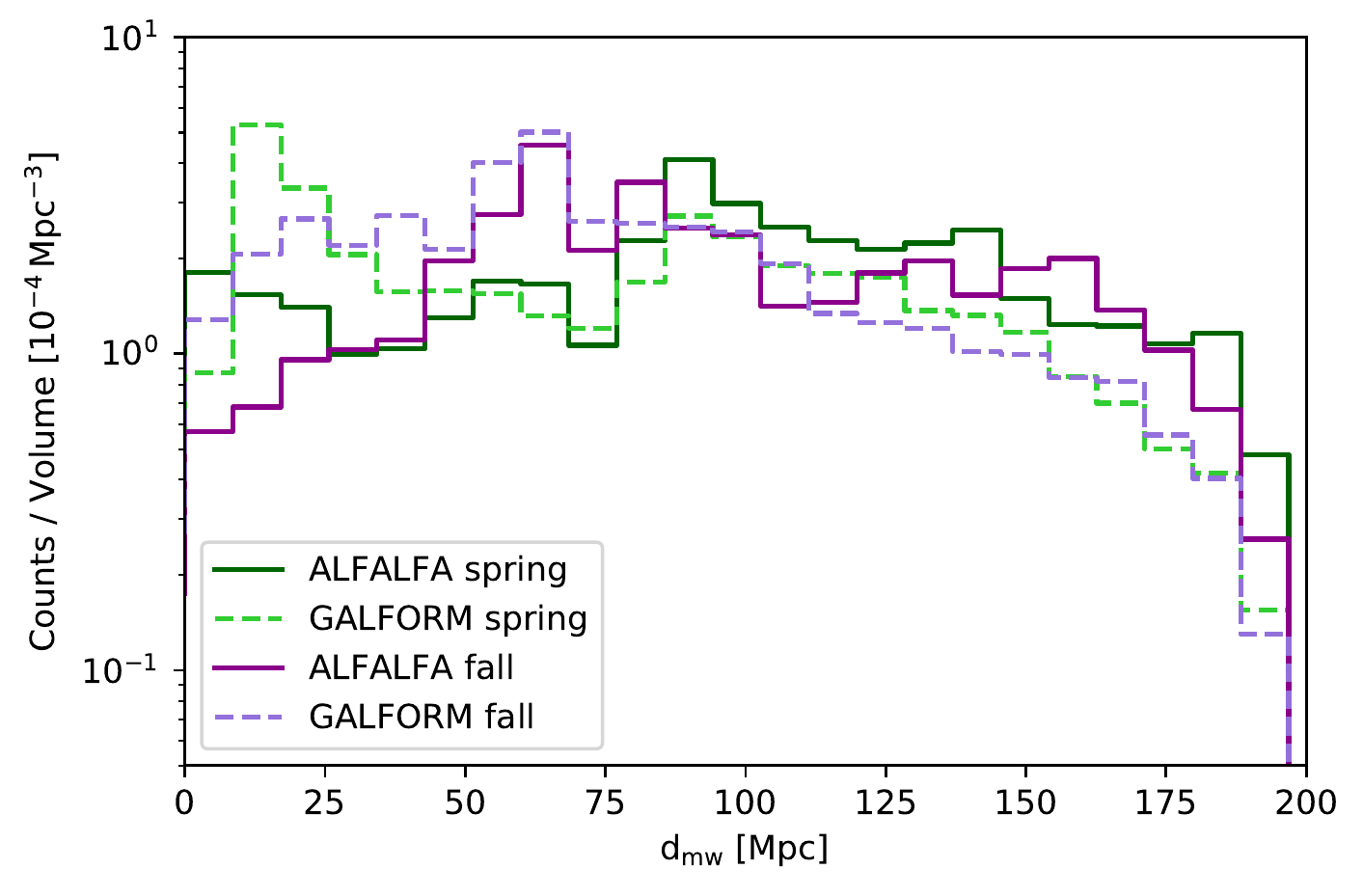}
    \caption{Histogram showing the counts per unit volume as a function of distance from the Milky Way, $d_{\mathrm{mw}}$. The $\alpha.100$ ALFALFA (solid lines) and \emph{GALFORM} (dashed lines) catalogues are shown, split by survey field: spring (ALFALFA: dark-green; \emph{GALFORM}: light-green) and fall (ALFALFA: magenta; \emph{GALFORM}: light purple).} 
    \label{fig:1D_distance_distribution}
\end{figure}

We have traced the origin of this feature to the difference in the number of detected sources as a function of distance (see Fig.~\ref{fig:1D_distance_distribution}) in the foreground of the survey ($d_{\mathrm{mw}}\lesssim 30\,\mathrm{Mpc}$). In Sec.~\ref{sec:LoSClustering} we confirmed that the systematic offset between the H\,\textsc{i}\,WFs in the two survey fields is not due to a global difference in the clustering of sources in distance. However, local features could still play a role. Essentially, the $1/\mathrm{V_{eff}}$ estimator seems to be incorrectly extrapolating the foreground overdensity in the spring field through the entire survey volume, biasing the spring field to higher number densities. This effect is similar to what would occur if we used the $1/V_\mathrm{max}$ estimator \citep{1968ApJ...151..393S}, which assumes that galaxies are uniformly distributed in space, but less severe -- the $1/V_{\mathrm{eff}}$ algorithm is intended to compensate for non-uniformity in the galaxy distribution, but does so imperfectly. The spring field in our mock \emph{GALFORM} survey also has slightly more sources than the fall field at small distances, but the effect is much less pronounced, explaining why this particular bias is less pronounced in the mock survey. This can be ascribed to the \emph{Sibelius-DARK} simulation being an imperfect match to the actual local cosmic web. We note an increased density of \emph{GALFORM} sources at $\sim15\,\mathrm{Mpc}$ in the spring field, associated with the Virgo cluster. We suggest that this is the cause of the localised overdense spring feature at $(M_{\mathrm{HI}},w_{50})\sim(10^{8.5}\,\mathrm{M}_\odot,120\,\mathrm{km}\,\mathrm{s}^{-1})$ in Fig.~\ref{fig:AA_GF_w50mHI_sf}. 

We confirm this interpretation of the origin of the offset between the ALFALFA spring and fall H\,\textsc{i}\,WFs by removing all sources from the ALFALFA catalogue with distances $d_{\mathrm{mw}} < 30\,\mathrm{Mpc}$. This choice of distance cut is further motivated in the lower panel of Fig.~\ref{fig:AA_HIWF_losclustering}. Here, we show the median ratio of the spring and fall H\,\textsc{i}\,WFs as a function of the minimum distance from the Milky~Way at which sources are retained. For all distances beyond about $30\,\mathrm{Mpc}$ the median ratio stays about the same, indicating that the foreground galaxies have the most influence within about this distance. In the middle panel of Fig.~\ref{fig:AA_HIWF_losclustering}, we show the H\,\textsc{i}\,WFs in the two fields measured with both (i)~the conservative CL described above imposed and (ii) sources within $30\,\mathrm{Mpc}$ removed. In this case the H\,\textsc{i}\,WFs in the two fields are very close to agreement; the median ratio between the spring to fall H\,\textsc{i}\,WFs derived using the independent $V_\mathrm{eff}$ weights is $\log_{10}(\phi_{s}\,/\,\phi_{f}) = 0.07 \pm 0.12$. We attribute the small remaining differences to the $1/V_{\mathrm{eff}}$ estimator slightly over-compensating for clustering of sources along the line of sight beyond $30\,\mathrm{Mpc}$. For comparison, we also show the H\,\textsc{i}\,WFs in the two fields measured using the $1/V_{\mathrm{max}}$ algorithm (i.e. assuming a spatially uniform galaxy distribution) in the middle sub-panel Fig.~\ref{fig:AA_HIWF_losclustering}. In this case, the sign of the offset between spring and fall is reversed (due to an overdensity of sources in the fall field at distances of $\sim50-80\,\mathrm{Mpc}$) with a median ratio between the spring to fall H\,\textsc{i}\,WFs derived using the independent $V_\mathrm{max}$ weights being $\log_{10}(\phi_{s}\,/\,\phi_{f}) = -0.01 \pm 0.12$.

In summary, we attribute the differences between the spring and fall ALFALFA H\,\textsc{i}\,WFs \citep[e.g. as measured by][]{2022MNRAS.509.3268O} to: (i)~the adopted completeness limit for the survey; and (ii)~the $1/V_\mathrm{eff}$ estimator incorrectly extrapolating the foreground overdensity in the spring field through the entire survey volume (in that field). Accounting for these systematic effects leads to H\,\textsc{i}\,WFs in the spring and fall fields that are consistent with being identical, which is in turn consistent with the two fields having identical HMFs (within a few per~cent) as expected in a $\Lambda$CDM cosmology.

\section{Discussion}\label{sec:Discussion} 

In Sec.~\ref{sec:systematics_CLs}, we comment further on the influence of the choice of CL on the H\,\textsc{i}\,WF and the importance for future surveys of mitigating variations in the CL. Then in Sec.~\ref{sec:Interpretation} we discuss the possible origins of the qualitative differences between the ALFALFA H\,\textsc{i}\,MF and H\,\textsc{i}\,WF and those that we obtain from simulations.

\subsection{Survey completeness limit}\label{sec:systematics_CLs}

For ALFALFA, the CL is determined empirically. \citet{2022MNRAS.509.3268O} found that the CLs in the ALFALFA spring and fall fields are different by about $0.02\,\mathrm{dex}$ in $S_{21}$ when derived separately. This small quantitative difference was assessed to be insufficient to have a strong influence on the shape (low-velocity slope, and location and sharpness of exponential cutoff at high velocities -- i.e. excluding the overall normalisation) of the H\,\textsc{i}\,WF. We agree with this assessment, but emphasise that the overall normalisation of the H\,\textsc{i}\,WF is extremely sensitive to the chosen CL. Given this, we have investigated whether differences between the spring and fall CLs could plausibly fully explain the overall difference in number density between the spring and fall H\,\textsc{i}\,WFs. Beginning from the global $50$~per~cent CL \citep[][eq.~A5]{2022MNRAS.509.3268O}, we gradually ‘raised' the fall and ‘lowered' the spring CLs by equal increments until the overall difference in number density in the two fields vanished. We found that the minimum offset between the CLs in order to fully account for the variation would have to be at least $0.1\,\mathrm{dex}$, or five times greater than the empirically determined difference. Such a large difference between the CLs is completely incompatible with the $\alpha.100$ catalogue. The difference between the CLs in the two fields can therefore only explain part of the overall difference in amplitude between the spring and fall ALFALFA H\,\textsc{i}\,WFs.

Since a difference in sensitivity between the spring and fall fields is apparent, it is worth considering whether there may also be variations in sensitivity internally within each field. Indeed, this has previously been measured: using a catalogue covering $40$~per~cent of the final surveyed area, \citet{2011AJ....142..170H} measured a CL that is $0.02\,\mathrm{dex}$ ‘deeper' than that measured by \citet{2022MNRAS.509.3268O} using the entire surveyed area in the spring region\footnote{This could, in principle, be attributed to differences between the algorithms used to derive the CLs, but \citet{2022MNRAS.509.3268O} re-derived a CL for the $\alpha.40$ catalogue (see their equations A2 \& A5) and found only much smaller differences with respect to the \citet{2011AJ....142..170H} measurement.}.

We suggest that it is worth considering possible measures to mitigate variations in sensitivity across the sky in the design of future {21-cm} surveys intending to measure the H\,\textsc{i}\,WF (or H\,\textsc{i}\,MF). As an illustrative example, a survey built up by covering a wide area to shallow depth and repeating this process to build up sensitivity might be expected to have a more spatially uniform CL in the presence of time-varying radio frequency interference than one where a small field is observed to a final target depth before moving on to a subsequent field.

We also comment briefly on the assumed form of the CL for ALFALFA: a two-segment broken power law. There is what we would characterise as tentative evidence in the ALFALFA catalogue data for departures from this form. For instance, there is a hint of an upturn at low line widths, and the sharp transition between the two power law segments may not accurately capture what is in reality likely to be a more gradual transition in slope. The relatively small number of sources observed in ALFALFA makes it challenging to make stronger statements, but the order of magnitude increase in source counts expected from the next generation of H\,\textsc{i} surveys should allow for much more tightly constrained determinations of their CLs.

\subsection{\texorpdfstring{Qualitative comparisons of observed and simulated H\,\textsc{i}\,MFs and H\,\textsc{i}\,WFs}{Qualitative comparisons of observed and simulated HIMFs and HIWFs}}\label{sec:Interpretation}

\subsubsection{The H\,\textsc{i}\,MF}\label{sec:Interpretation_GFHIMF}

The true \emph{GALFORM} H\,\textsc{i}\,MF does not reproduce the almost constant low-mass slope seen for ALFALFA (left panels, Fig.~\ref{fig:HIMF_HIWF}). Instead, there is a fall and rise in the number density of sources over the mass range $10^{6}<M_{\mathrm{HI}}/\mathrm{M}_{\odot}< 10^{8}$. This is caused by the relative contributions of central and satellite galaxies to the H\,\textsc{i}\,MF, which in \emph{GALFORM} are easily separable. In \emph{GALFORM}, all galaxies are assumed to originate as central galaxies, but when the dark matter halo of a galaxy merges\footnote{As defined by \citet{2016MNRAS.462.3854L}.} with another more massive halo, that galaxy is assumed to become a satellite in the new larger halo. The evolution of galaxy properties is modelled separately for centrals and satellites. For instance, as soon as a galaxy becomes a satellite, its hot gas halo is instantly stripped away by the ram pressure of the central galaxy, and added to the central's hot gas halo. Consequently, no further gas is allowed to cool and accrete onto the satellite. This aggressive stripping influences the cold gas content of a galaxy, causing the dip in the distribution in H\,{\textsc{i}} mass, seen in our Fig.~\ref{fig:HIMF_HIWF} to occur for $M_\mathrm{HI}\lesssim10^{8}\,\mathrm{M}_{\odot}$. The exact location of this transition is affected by the halo mass resolution of the simulation \citep[][Sec.~4.2.2]{2016MNRAS.462.3854L}. However, the dominance of the H\,\textsc{i}\,MF by central and satellite galaxies in the high- and low-mass ends, respectively, is independent of the model adopted \citep[][Sec.~4.1.1]{2011MNRAS.418.1649L}. Smoothing these two peaks into a constant low-mass slope for the H\,\textsc{i}\,MF may be possible if satellites' hot gas is not instantaneously ram pressure stripped and/or gas is allowed to cool onto satellites.

\subsubsection{The H\,\textsc{i}\,WF}\label{sec:Interpretation_GF_HIWF}

The true \emph{GALFORM} H\,\textsc{i}\,WF when compared to ALFALFA shows a distinct lack of larger line width sources, particularly for $w_{50} > 200\,\mathrm{km}\,\mathrm{s}^{-1}$ (Fig.~\ref{fig:HIMF_HIWF}, right panels). The \citet{2016MNRAS.462.3854L} \emph{GALFORM} model assumes that atomic gas is distributed in a relatively compact disc (with the same scale-length as the stellar disc) in all galaxies. Cases where the emission is dominated instead by a much more diffuse disc, a ring, or even discrete clouds, such as is observed in some massive galaxies \citep[e.g.][]{1999PASA...16...28O,2012MNRAS.422.1835S}, are not modelled. This leads to severe underestimates of $w_{50}$ for some galaxies where a more realistic, radially extended H\,\textsc{i} distribution would lead to a wider line. The deficit of galaxies at the highest H\,\textsc{i} masses (visible in the H\,\textsc{i}\,MF), which should be populated by galaxies with large total masses and correspondingly large H\,\textsc{i} line widths, likely also plays a role.

\section{\texorpdfstring{Summary and outlook -- the H\,\textsc{i}\,WF as a cosmological constraint}{Summary and outlook -- the HIWF as a cosmological constraint}}\label{sec:ConstraintDM_conc}

The low line width end of the H\,\textsc{i}\,WF is in principle sensitive to properties of the dark matter through its close relationship with the HMF. Our H\,\textsc{i}\,WF derived from the $\Lambda$CDM \emph{Sibelius} simulations (Fig.~\ref{fig:HIMF_HIWF}) has a much shallower slope than expected from the HMF in the mock-surveyed volume, and is in reasonable qualitative agreement with that measured from ALFALFA observations at low line-widths. This reinforces previous work showing that the ALFALFA and HIPASS H\,\textsc{i}\,WFs can be reconciled with a $\Lambda$CDM cosmology once relevant galaxy formation processes and observational biases are accounted for \citep{2013MNRAS.431.1366S,2017ApJ...850...97B, 2019MNRAS.482.5606D}. Specifically, this is a consequence of: (i) not every low-mass \emph{Sibelius-DARK} halo hosting a \emph{GALFORM} galaxy; (ii) not every galaxy having substantial amounts of atomic gas ($M_{\mathrm{HI}}>10^{6}\,\mathrm{M}_{\odot}$); and (iii) the non-linear mapping between the maximum circular velocity and $w_{50}$. We have not, however, ruled out interpretations involving a low-mass truncation of the HMF, such as may arise from a WDM particle \citep[e.g.][]{2009ApJ...700.1779Z}. We plan to apply our mock survey methodology to a WDM version of the \emph{Sibelius-DARK} volume in future work.

Our approach using a constrained N-body realisation of the entire ALFALFA survey volume has allowed us to investigate systematic effects inaccessible to previous studies. The \emph{Sibelius-DARK} simulation makes the clear prediction that the spring and fall fields of the survey should have identical HMFs, and our modelling extends this to predicting identical H\,\textsc{i}\,MFs and H\,\textsc{i}\,WFs. Taken at face value, this is in tension with the observed differences between the spring and fall H\,\textsc{i}\,MFs and H\,\textsc{i}\,WFs measured from ALFALFA observations \citep{2018MNRAS.477....2J, 2022MNRAS.509.3268O}. We identified two effects that can bias the H\,\textsc{i}\,WF and plausibly explain the observed differences between the spring and fall ALFALFA H\,\textsc{i}\,WFs:
\begin{enumerate}
\item The CLs in the spring and fall survey fields differ, but have previously been assumed to be identical in deriving the $1/V_\mathrm{eff}$ weights used in deriving the H\,\textsc{i}\,WF. We tabulate updated measurements correcting for this in Appendix~\ref{appendix:tables}.
\item The $1/V_{\mathrm{eff}}$ algorithm is intended to compensate for galaxy clustering (c.f. the classical $1/V_{\mathrm{max}}$ method assuming a uniform spatial distribution), but does so imperfectly. A foreground over-density of H\,\textsc{i}-rich galaxies in the spring ALFALFA field, and an underdensity in the foreground of the fall field, drive large systematic errors.
\end{enumerate}
We note that the small remaining systematic offset between the spring and fall H\,\textsc{i}\,WFs once the above two effects are accounted for is most likely also attributable to the limited ability of the $1/V_{\mathrm{eff}}$ to accurately compensate for differences between the galaxy clustering in distance in the two fields. In this interpretation, the true H\,\textsc{i}\,WF is spatially invariant (within about 10~per~cent) when sampled in volumes comparable to the volume of the ALFALFA survey; the observed differences between the spring and fall H\,\textsc{i}\,WFs in ALFALFA are entirely spurious. Encouragingly, the influence of individual over/underdense regions on the calculation of $1/V_\mathrm{eff}$ weights can be mitigated by simply surveying a larger area on the sky; the ongoing WALLABY\footnote{Widefield ASKAP L-band Legacy All-sky Blind surveY} survey \citep{2020Ap&SS.365..118K} will cover an area about four times wider than that covered by ALFALFA.

The H\,\textsc{i}\,WF has the potential to become a stringent test of cosmological models. Realising this potential will require a deeper understanding of the systematic biases influencing measurements, such as those due to spatial (or temporal) variability in survey sensitivity. Progress on theoretical issues is also needed. Combining semi-analytical models able to simulate directly the large volumes of current and future surveys with hydrodynamical simulations able to resolve in detail the internal structure of atomic gas in galaxies seems a promising path forward. We look forward to the prospect of using the H\,\textsc{i}\,WF as a constraint on the nature of dark matter.

\section*{Acknowledgements}

We thank the anonymous referee for a careful review of the manuscript that led to numerous improvements. KAO and CSF acknowledge support by the European Research Council (ERC) through Advanced Investigator grant to C.S. Frenk, DMIDAS (GA 786910). KAO acknowledges support by STFC through grant ST/T000244/1. RANB acknowledges support from both Durham University's Collingwood College Undergraduate Research Internships (UGRI) programme and the Royal Society. This work used the DiRAC@Durham facility managed by the Institute for Computational Cosmology on behalf of the STFC DiRAC HPC Facility (www.dirac.ac.uk). The equipment was funded by BEIS capital funding via STFC capital grants ST/K00042X/1, ST/P002293/1, ST/R002371/1 and ST/S002502/1, Durham University and STFC operations grant ST/R000832/1. DiRAC is part of the National e-Infrastructure. This research has made use of NASA’s Astrophysics Data System.

\section*{Data Availability}
 
The public release of the \emph{Sibelius-DARK} + \emph{GALFORM} simulation data is available from \url{https://virgodb.dur.ac.uk/} as described in \cite[][appendix~A]{2022MNRAS.512.5823M}. Fully reproducing our analysis requires some \emph{GALFORM} galaxy properties not included in the public release (e.g. bulge fractions, stellar half-mass radii, etc.). These can be requested from the Sibelius project team \citep{2022MNRAS.512.5823M}. The $\alpha.100$ catalogue is available from: \url{http://egg.astro.cornell.edu/alfalfa/data/index.php}.

\bibliographystyle{mnras}
\bibliography{reflist}

\appendix

\section{Completeness of the ALFALFA catalogues}
\label{app:CLs}

We present the $\alpha.100$ catalogue CL for the spring and fall fields individually. The global CL was measured and given in \citet[][equations~A4--A6]{2022MNRAS.509.3268O}. The fall CL is slightly shallower than the global CL, by $0.011 \,\mathrm{dex}$, while the spring CL is slightly deeper, by $0.009 \,\mathrm{dex}$, for a net difference of $0.02\,\mathrm{dex}$. Using the $\alpha.100$ catalogue for spring sources only, we derive the following CLs:

\begin{equation} \label{90_CL_S}
    \log_{10}\left(\frac{S_{21,90\%}}{\mathrm{Jy}\,\mathrm{km}\,\mathrm{s}^{-1}}\right) =
    \begin{cases}
      0.5\,W - 1.124 & \text{$W < 2.5$}\\
      W - 2.374 & \text{$W \geq 2.5$}\\
    \end{cases}
\end{equation}

\begin{equation} \label{50_CL_S_app}
    \log_{10}\left(\frac{S_{21,50\%}}{\mathrm{Jy}\,\mathrm{km}\,\mathrm{s}^{-1}}\right) =
    \begin{cases}
      0.5\,W - 1.179 & \text{$W < 2.5$}\\
      W - 2.429 & \text{$W \geq 2.5$}\\
    \end{cases}
\end{equation}

\begin{equation} \label{25_CL_S}
    \log_{10}\left(\frac{S_{21,25\%}}{\mathrm{Jy}\,\mathrm{km}\,\mathrm{s}^{-1}}\right) =
    \begin{cases}
      0.5\,W - 1.207 & \text{$W < 2.5$}\\
      W - 2.457 & \text{$W \geq 2.5$}\\
    \end{cases}
\end{equation}

\noindent
where $W = \log_{10}(w_{50}\,/\mathrm{km}\,\mathrm{s}^{-1})$. Similarly, using the $\alpha.100$ catalogue for fall sources only, we derive the following CLs:

\begin{equation} \label{90_CL_F}
    \log_{10}\left(\frac{S_{21,90\%}}{\mathrm{Jy}\,\mathrm{km}\,\mathrm{s}^{-1}}\right) =
    \begin{cases}
      0.5\,W - 1.104 & \text{$W < 2.5$}\\
      W - 2.354 & \text{$W \geq 2.5$}\\
    \end{cases}
\end{equation}

\begin{equation} \label{50_CL_F_app}
    \log_{10}\left(\frac{S_{21,50\%}}{\mathrm{Jy}\,\mathrm{km}\,\mathrm{s}^{-1}}\right) =
    \begin{cases}
      0.5\,W - 1.159 & \text{$W < 2.5$}\\
      W - 2.409 & \text{$W \geq 2.5$}\\
    \end{cases}
\end{equation}

\begin{equation} \label{25_CL_F}
    \log_{10}\left(\frac{S_{21,25\%}}{\mathrm{Jy}\,\mathrm{km}\,\mathrm{s}^{-1}}\right) =
    \begin{cases}
      0.5\,W - 1.187 & \text{$W < 2.5$}\\
      W - 2.237 & \text{$W \geq 2.5$}\\
    \end{cases}
\end{equation}

\section{\emph{Sibelius-DARK} Octant Data}\label{appendix:octants}

We present in Table~\ref{table:octants} the vertices in right ascension and declination for the octants of the \emph{Sibelius-DARK} sky used in Fig.~\ref{fig:X}.

\begin{table*}
\centering
\caption{The sky coordinate position definition and number counts of all dark matter haloes contained within bounds of each \emph{Sibelius-DARK} octant. The median octant contains 2.42 million dark matter haloes. A distance cut keeping only sources within $\mathrm{d_{mw} \leq 200\,\mathrm{Mpc}}$ is applied.}
\begin{tabular}{|c|c|c|}
 \hline
 \multicolumn{3}{|c|}{}\\$\mathrm{Right\,Ascension}\,[^{\circ}]$& $\mathrm{Declination}\,[^{\circ}]$ & Number of sources [$\times10^{6}$]  \\
 \hline
 0 - 90 &  0 - 90 & 2.61\\
0 - 90 &  $-90$ - 0 & 2.55\\
90 - 180 &  0 - 90 & 2.30\\
90 - 180 &  $-90$ - 0 & 2.62\\
180 - 270 &  0 - 90 & 2.73\\
180 - 270 &  $-90$ - 0 & 2.40\\
270 - 360& 0 - 90 & 1.93\\
270 - 360&  $-90$ - 0 & 2.04\\
 \hline
 \label{table:octants}
\end{tabular}
\end{table*}

\section{\texorpdfstring{Tabulated ALFALFA H\,{\textsc{i}}\,WF and H\,{\textsc{i}}\,MF}{Tabulated ALFALFA HIWF and HIMF}}\label{appendix:tables}

In Tables \ref{table:New_AA_HIWF} and \ref{table:new_AA_HIMF} we tabulate our $\alpha.100$ ALFALFA H\,{\textsc{i}}\,WF and H\,{\textsc{i}}\,MF measurements; see bottom panels of Fig.~\ref{fig:HIMF_HIWF}.

\begin{table*}
\centering
\caption{Amplitudes and uncertainties of the measured $\alpha.100$ H\,{\textsc{i}}\,WF for the spring and fall fields derived assuming the completeness limits given in Equations~(\ref{90_CL_S}--\ref{25_CL_F}) and a cut in recessional velocity, $v_{\mathrm{rec}} \leq 15,000\,\mathrm{km}\,\mathrm{s}^{-1}$, matching that used in \citet{2022MNRAS.509.3268O}. The combined measurement for the entire survey uses the $1/V_{\mathrm{eff}}$ weights calculated separately for the two fields. Note that although the logarithms of all amplitudes and uncertainties are given, the uncertainties are symmetric on a linear scale and should be interpreted as the $1\sigma$ width of a Gaussian distribution (not a log-normal distribution).}
\label{tab:my_label2}
\begin{tabular}{|c|c|c|c|}
 \hline
 \multicolumn{4}{|c|}{$\log_{10}\phi(w_{50})\,/\mathrm{Mpc}^{-3}\,\mathrm{dex}^{-1}$} \\
 \hline$\log_{10}\,w_{50}\,/\mathrm{km}\,\mathrm{s}^{-1}$ & $\alpha.100$ Spring & $\alpha.100$ Fall & $\alpha.100$\\
 \hline
 1.25& $-0.35^{+0.18}_{-0.32}$ & $-1.36^{+0.24}_{-0.59}$ & $-0.53^{+0.18}_{-0.3}$ \\
 1.35& $-0.40^{+0.13}_{-0.18}$ & $-0.71^{+0.15}_{-0.24}$ & $-0.49^{+0.11}_{-0.14}$\\
 1.45& $-0.82^{+0.06}_{-0.08}$ & $-0.70^{+0.14}_{-0.20}$ & $-0.77^{+0.07}_{-0.09}$\\
 1.55& $-0.66^{+0.07}_{-0.08}$ & $-0.88^{+0.09}_{-0.11}$ & $-0.74^{+0.06}_{-0.07}$\\
 1.65& $-0.82^{+0.05}_{-0.06}$ & $-1.10^{+0.09}_{-0.11}$ & $-0.89^{+0.04}_{-0.05}$\\
 1.75& $-0.92^{+0.05}_{-0.05}$ & $-1.07^{+0.07}_{-0.08}$ & $-0.97^{+0.04}_{-0.04}$\\
 1.85& $-0.90^{+0.06}_{-0.06}$ & $-1.20^{+0.06}_{-0.07}$ & $-0.10^{+0.05}_{-0.05}$\\
 1.95& $-0.82^{+0.05}_{-0.05}$ & $-1.22^{+0.05}_{-0.05}$ & $-0.93^{+0.04}_{-0.04}$\\
 2.05& $-0.91^{+0.05}_{-0.06}$ & $-1.25^{+0.06}_{-0.07}$ & $-1.03^{+0.04}_{-0.05}$\\
 2.15& $-1.06^{+0.04}_{-0.05}$ & $-1.42^{+0.03}_{-0.04}$ & $-1.14^{+0.04}_{-0.04}$\\
 2.25& $-1.22^{+0.03}_{-0.03}$ & $-1.45^{+0.04}_{-0.04}$ & $-1.29^{+0.03}_{-0.03}$\\
 2.35& $-1.49^{+0.03}_{-0.03}$ & $-1.69^{+0.03}_{-0.03}$ & $-1.54^{+0.02}_{-0.02}$\\
 2.45& $-1.57^{+0.05}_{-0.05}$ & $-1.87^{+0.03}_{-0.04}$ & $-1.66^{+0.04}_{-0.04}$\\
 2.55& $-1.89^{+0.04}_{-0.05}$ & $-2.09^{+0.06}_{-0.07}$ & $-1.94^{+0.04}_{-0.04}$\\
 2.65& $-2.02^{+0.09}_{-0.11}$ & $-2.47^{+0.06}_{-0.07}$ & $-2.12^{+0.07}_{-0.08}$\\
 2.75& $-2.90^{+0.08}_{-0.10}$ & $-3.07^{+0.06}_{-0.07}$ & $-2.87^{+0.05}_{-0.06}$\\
 2.85& $-4.07^{+0.15}_{-0.23}$ & $-4.42^{+0.16}_{-0.25}$ & $-4.08^{+0.11}_{-0.15}$\\
 \hline
 \label{table:New_AA_HIWF}
\end{tabular}
\end{table*}

\begin{table*}
\centering
\caption{As Table \ref{table:New_AA_HIWF} but for the H\,{\textsc{i}}\,MF.}
\label{tab:my_label3}
\begin{tabular}{|c|c|c|c|}
 \hline
 \multicolumn{4}{|c|}{$\log_{10}\phi(M_{\mathrm{HI}})\,/\mathrm{Mpc}^{-3}\,\mathrm{dex}^{-1}$} \\
 \hline$\log_{10}\,M_{\mathrm{HI}}\,/\mathrm{M}_{\odot}$ & $\alpha.100$ Spring & $\alpha.100$ Fall & $\alpha.100$ \\
 \hline
  6.1& $-0.80^{+0.23}_{-0.53}$ & - & $-1.01^{+0.23}_{-0.53}$\\
 6.3& $-1.28^{+0.30}_{-\infty}$ & - & $-1.49^{+0.30}_{-1.57}$\\
 6.5& $-1.09^{+0.20}_{-0.38}$ & $-1.52^{+0.30}_{-\infty}$ & $-1.20^{+0.18}_{-0.31}$\\
 6.7& $-1.22^{+0.17}_{-0.27}$ & $-1.17^{+0.20}_{-0.39}$ & $-1.20^{+0.14}_{-0.20}$\\
 6.9& $-1.49^{+0.13}_{-0.18}$ & $-1.21^{+0.15}_{-0.23}$ & $-1.36^{+0.10}_{-0.14}$\\
 7.1& $-1.01^{+0.07}_{-0.08}$ & $-1.47^{+0.14}_{-0.21}$ & $-1.14^{+0.06}_{-0.07}$\\
 7.3& $-0.96^{+0.06}_{-0.07}$ & $-1.49^{+0.12}_{-0.16}$ & $-1.10^{+0.05}_{-0.06}$\\
 7.5& $-1.03^{+0.05}_{-0.05}$ & $-1.41^{+0.10}_{-0.12}$ & $-1.14^{+0.04}_{-0.05}$\\
 7.7& $-1.19^{+0.04}_{-0.04}$ & $-1.38^{+0.08}_{-0.09}$ & $-1.25^{+0.04}_{-0.04}$\\
 7.9& $-1.28^{+0.04}_{-0.04}$ & $-1.43^{+0.07}_{-0.08}$ & $-1.33^{+0.03}_{-0.04}$\\
 8.1& $-1.33^{+0.03}_{-0.03}$ & $-1.52^{+0.06}_{-0.07}$ & $-1.40^{+0.03}_{-0.03}$\\
 8.3& $-1.34^{+0.03}_{-0.03}$ & $-1.66^{+0.05}_{-0.06}$ & $-1.44^{+0.02}_{-0.03}$\\
 8.5& $-1.42^{+0.03}_{-0.03}$ & $-1.64^{+0.04}_{-0.04}$ & $-1.49^{+0.02}_{-0.02}$\\
 8.7& $-1.49^{+0.02}_{-0.02}$ & $-1.66^{+0.03}_{-0.04}$ & $-1.55^{+0.02}_{-0.02}$\\
 8.9& $-1.58^{+0.02}_{-0.02}$ & $-1.72^{+0.02}_{-0.03}$ & $-1.63^{+0.01}_{-0.02}$\\
 9.1& $-1.63^{+0.02}_{-0.02}$ & $-1.74^{+0.02}_{-0.02}$ & $-1.67^{+0.01}_{-0.01}$\\
 9.3& $-1.75^{+0.01}_{-0.01}$ & $-1.86^{+0.01}_{-0.01}$ & $-1.79^{+0.01}_{-0.01}$\\
 9.5& $-1.92^{+0.01}_{-0.01}$ & $-2.01^{+0.01}_{-0.01}$ & $-1.95^{+0.01}_{-0.01}$\\
 9.7& $-2.09^{+0.01}_{-0.01}$ & $-2.16^{+0.01}_{-0.01}$ & $-2.11^{+0.01}_{-0.01}$\\
 9.9& $-2.23^{+0.01}_{-0.01}$ & $-2.38^{+0.01}_{-0.01}$ & $-2.32^{+0.01}_{-0.01}$\\
 10.1& $-2.59^{+0.01}_{-0.01}$ & $-2.67^{+0.01}_{-0.01}$ & $-2.62^{+0.01}_{-0.01}$\\
 10.3& $-3.06^{+0.02}_{-0.02}$ & $-3.09^{+0.02}_{-0.02}$ & $-3.07^{+0.01}_{-0.01}$\\
 10.5& $-3.74^{+0.03}_{-0.04}$ & $-3.83^{+0.05}_{-0.05}$ & $-3.77^{+0.03}_{-0.03}$\\
 10.7& $-4.62^{+0.09}_{-0.11}$ & $-4.65^{+0.11}_{-0.16}$ & $-4.64^{+0.07}_{-0.09}$\\
 10.9& $-5.61^{+0.23}_{-0.53}$ & - & $-5.81^{+0.23}_{-0.53}$\\
 \hline
 \label{table:new_AA_HIMF}
\end{tabular}
\end{table*}

\bsp
\label{lastpage}
\end{document}